\documentclass[twocolumn,iop,numberedappendix]{emulateapj}

\usepackage{graphics,color}
\graphicspath{{Figures/}{./}}

\usepackage{CJKutf8}

\newcommand{\code}[1]{\texttt{#1}}

\begin{document}

\title{The Eleventh and Twelfth Data Releases of the Sloan Digital Sky Survey:  Final Data from SDSS-III}

\author{
\begin{CJK*}{UTF8}{gbsn}
Shadab~Alam\altaffilmark{1},
Franco~D.~Albareti\altaffilmark{2},
Carlos~{Allende~Prieto}\altaffilmark{3,4},
F.~Anders\altaffilmark{5},
Scott~F.~Anderson\altaffilmark{6},
Timothy~Anderton\altaffilmark{7},
Brett~H.~Andrews\altaffilmark{8,9},
Eric~Armengaud\altaffilmark{10},
\'Eric~Aubourg\altaffilmark{11},
Stephen~Bailey\altaffilmark{12}
% ...
\end{CJK*}
}

\altaffiltext{1}{
McWilliams Center for Cosmology,
Department of Physics, 
Carnegie Mellon University, 5000 Forbes Ave, Pittsburgh, PA 15213, USA
}

\altaffiltext{2}{
Instituto de F\'{\i}sica Te\'orica, (UAM/CSIC), 
Universidad Aut\'onoma de Madrid, Cantoblanco, E-28049 Madrid, Spain
}

\altaffiltext{3}{
Instituto de Astrof{\'\i}sica de Canarias (IAC), 
C/V{\'\i}a L\'actea,
s/n, E-38200, La Laguna, Tenerife, Spain
}

\altaffiltext{4}{
Departamento de Astrof\'{\i}sica, 
Universidad de La Laguna, 
E-38206, La Laguna, Tenerife, Spain
}

\altaffiltext{5}{
Leibniz-Institut f\"ur Astrophysik Potsdam (AIP), 
An der Sternwarte 16, 
D-14482 Potsdam, Germany
}

\altaffiltext{6}{
Department of Astronomy, University of Washington, 
Box 351580, Seattle, WA 98195, USA
}

\altaffiltext{7}{
Department of Physics and Astronomy, 
University of Utah, Salt Lake City, UT 84112, USA
}

\altaffiltext{8}{
Department of Astronomy, 
Ohio State University, 140 West 18th Avenue, Columbus, OH 43210, USA
}

\altaffiltext{9}{
PITT PACC, Department of Physics and Astronomy, 
University of Pittsburgh, 
3941 O'Hara Street,
Pittsburgh, PA 15260, USA
}

\altaffiltext{10}{
CEA, Centre de Saclay, Irfu/SPP,  F-91191 Gif-sur-Yvette, France
}

\altaffiltext{11}{
APC, University of Paris Diderot, CNRS/IN2P3, CEA/IRFU, Observatoire de Paris, Sorbonne Paris Cit\'e, F-75205 Paris, France
}

\altaffiltext{12}{
Lawrence Berkeley National Laboratory, One Cyclotron Road,
Berkeley, CA 94720, USA
}

% ...

\shorttitle{SDSS DR12}

\begin{abstract}
The third generation of the Sloan Digital Sky Survey (\mbox{SDSS-III}) took data
from 2008 to 2014
using the original SDSS wide-field imager, the original and an
upgraded multi-object fiber-fed optical spectrograph, a new
near-infrared high-resolution spectrograph, and a novel optical interferometer.
All the data from \mbox{SDSS-III} are now made public.  In particular, this
paper describes Data Release 11 (DR11) including all data acquired
through 2013 July, and 
Data Release 12 (DR12) adding data acquired through 2014 July
(including all data included in previous data releases),
marking the end of \mbox{SDSS-III} observing.
Relative to our previous public release (DR10), DR12 adds one million
new spectra of galaxies and quasars from the Baryon Oscillation
Spectroscopic Survey (BOSS) over an additional 3000~deg$^2$ of sky, 
 more than triples the number of $H$-band spectra of
stars as part of the Apache Point Observatory (APO) Galactic Evolution
Experiment (APOGEE), and includes repeated accurate radial velocity
measurements of 5500 stars from the Multi-Object APO Radial Velocity
Exoplanet Large-area Survey (MARVELS).  The APOGEE outputs now include
measured abundances of 15 different elements for each star.  
In total, \mbox{SDSS-III} added 2350 deg$^2$ of $ugriz$ imaging; 155,520
spectra of 138,099 stars as part of the Sloan Exploration of Galactic
Understanding and Evolution 2 (SEGUE-2) survey; 2,497,484 BOSS spectra
of 1,372,737 galaxies, 294,512 quasars, and 247,216 stars over 9376
deg$^2$; 618,080 APOGEE spectra of 156,593 stars; and 197,040 MARVELS
spectra of 5,513 stars. Since its first light in 1998, SDSS has imaged
over 1/3 of the Celestial sphere in five bands and obtained over five
million astronomical spectra.

\end{abstract}

\keywords{Atlases---Catalogs---Surveys}

\section{Introduction}
\label{sec:introduction}

Comprehensive wide-field imaging and spectroscopic surveys of the sky
have played a key role in astronomy, leading to fundamental new
breakthroughs in our understanding of the Solar System;
 our Milky Way Galaxy and its constituent stars and gas;
 the nature, properties, and evolution of galaxies;
 and the Universe as a whole.  
The Sloan Digital Sky Survey (SDSS), which started routine operations in 2000
April, has carried out imaging and spectroscopy over roughly 1/3 of
the Celestial Sphere.  The SDSS uses a dedicated 2.5-meter wide-field telescope
\citep{Gunn06}, instrumented with a sequence of sophisticated imagers
and spectrographs.  The SDSS has gone through a series of stages.
\mbox{SDSS-I} \citep{York00}, which was in operation through 2005, focused on
a ``Legacy'' survey of five-band imaging (using what was at the time the largest
camera ever used in optical astronomy; \citealt{Gunn98}) and
spectroscopy of well-defined samples of galaxies
\citep{Strauss02,Eisenstein01} and quasars \citep{Richards02}, using a
640-fiber pair of spectrographs \citep{Smee13}.  \mbox{SDSS-II}
operated from 2005 to 2008, and finished the Legacy survey.  It also 
carried out a repeated imaging survey of the Celestial Equator in 
the Fall sky to search for supernovae \citep{Frieman08}, as well as a
spectroscopic survey of stars to study the structure of the Milky Way
\citep{Yanny09}. 

\mbox{SDSS-III} \citep{Eisenstein11} started operations in Fall 2008,
completing in Summer 2014.  \mbox{SDSS-III} consisted of four interlocking surveys: 
\begin{itemize} 
\item The {\bf Sloan Exploration of Galactic Understanding and
  Evolution 2} (SEGUE-2; C.~Rockosi et al.~2015, in preparation) used the \mbox{SDSS-I/II} spectrographs to obtain
  $R\sim 2000$ spectra of 
stars at high and low
  Galactic latitudes to study Galactic structure, dynamics, and
  stellar populations. SEGUE-2 gathered data during the 2008--2009 season.
\item The {\bf Baryon Oscillation Spectroscopic Survey} (BOSS; \citealt{Dawson13}) used
  the SDSS imager to increase the footprint of the SDSS imaging in the
  Southern Galactic Cap in the 2008--2009 season.  The SDSS
  spectrographs were then completely rebuilt, with new fibers ($2''$
  entrance aperture rather than $3''$, 1000 fibers per exposure), as
  well as new gratings, CCDs, and optics. 
  Galaxies (B.~Reid et al.~2015, in preparation) 
  and quasars \citep{Ross12} were
  selected from the SDSS imaging data, and are used to study the
  baryon oscillation feature in the clustering of galaxies
  \citep{Anderson14a,Anderson14b} and Lyman-$\alpha$ absorption along the line of sight to distant quasars \citep{Busca13,Slosar13,Font-Ribera14,Delubac15}.  
  BOSS collected spectroscopic data from 2009 December to 2014 July.
\item The {\bf Apache Point Observatory Galaxy Evolution Experiment}
  (APOGEE; S.~Majewski et al.~2015, in preparation) used a separate 300-fiber
  high-resolution ($R \sim 22,500$) $H$-band spectrograph
  to investigate the composition and dynamics of stars in the Galaxy.
  The target
  stars were selected from the database of the Two Micron All-Sky Survey
  (2MASS; \citealt{Skrutskie06}); the resulting spectra give highly
  accurate stellar surface temperatures, gravities, and detailed
  abundance measurements.  APOGEE gathered data from 2011 May to 2014 July.
\item The {\bf Multi-Object APO
Radial Velocity Exoplanet Large-area Survey} 
  (MARVELS; J.~Ge et al.~2015, in preparation) used a 60-fiber
  interferometric spectrograph to measure high-precision radial
  velocities of stars to search for extra-solar planets and brown
  dwarfs orbiting them. 
  MARVELS gathered data from 2008 October to 2012 July.
\end{itemize}

The SDSS data have been made available to the scientific community and
the public in a roughly annual cumulative series of data releases.
These data have been distributed~\citep{Thakar2008a} in the form of direct access to raw
and processed imaging and spectral files and also through a relational
database (the ``Catalog Archive Server'', or ``CAS''), presenting the
derived catalog information.  As of DR12 these 
catalogs present information on a total of $\sim$470 million objects
in the imaging survey, and 5.3 million spectra. 

  The Early Data
Release (EDR; \citealt{EDR}), and Data Releases 1--5 
(DR1; \citealt{DR1}, 
DR2; \citealt{DR2},
DR3; \citealt{DR3},
DR4; \citealt{DR4}, and DR5; \citealt{DR5}) included data from SDSS-I.
DR6 and DR7 \citep{DR6,DR7} covered the data in SDSS-II. The data from
\mbox{SDSS-III} have appeared in three releases thus far.  DR8~\citep{DR8}
included the final data from the SDSS imaging camera, as well as all
the SEGUE-2 data.  DR9~\citep{DR9} included the first spectroscopic
data from BOSS.  DR10~\citep{DR10} roughly doubled the amount of BOSS
data made public, and included the first release of APOGEE data. 

The \mbox{SDSS-III} collaboration has found it useful to internally define a
data set associated with the data taken through 2013 Summer, which we
designate as ``DR11''.  The \mbox{SDSS-III} completed data-taking in 2014 July, and
the present paper describes both DR11 and Data Release 12 (DR12).
Like previous data releases, DR12 is cumulative; it
includes all data taken by SDSS to date.  
DR12 includes almost 2.5 million BOSS spectra
of quasars, galaxies, and stars over 9,376 square degrees:
155,000 SEGUE-2 spectra of 138,000 stars (as released in DR8),
and 618,000 APOGEE spectra of 156,000 stars. 
It also includes the
first release of MARVELS data, presenting 197,000 spectra of 5,500
stars (3,300 stars with $>16$ observations each).
Because some BOSS, APOGEE, and MARVELS
scientific papers have been based on the DR11 sample, this paper describes the
distinction between DR11 and DR12 and the processing software for the
two data sets, and how to understand
this distinction in the database.  

The data release itself may be accessed from the SDSS-III
website\footnote{\url{http://www.sdss3.org/dr12}} or the DR12 page of
the new pan-SDSS website.\footnote{\url{http://www.sdss.org/dr12}}  
DR11 is similarly available through the same interfaces.
The outline of this 
paper is as follows.  We summarize the full contents of DR11 and DR12
in Section~\ref{sec:coverage}, emphasizing the quantity of spectra and the
solid angle covered by each of the surveys.  Details for each
component of \mbox{SDSS-III} are described in Section~\ref{sec:marvels} (MARVELS), Section~\ref{sec:boss} (BOSS) and
Section~\ref{sec:apogee} (APOGEE).  There have been no updates to SEGUE-2
since DR9 and we do not discuss it further in this
paper.  We describe the distribution of the data 
in Section~\ref{sec:distribution}, and conclude, with a view to
the future, in Section~\ref{sec:future}. 

\section{Summary of coverage}
\label{sec:coverage}

DR12 presents all data gathered by SDSS-III, which extended from 2008
August to 2014 June, plus a small amount of data gathered with the
BOSS and APOGEE instruments in the first two weeks of 2014 July under
the auspices of the next phase of the Sloan Digital Sky Survey,
\mbox{SDSS-IV} (see Section~\ref{sec:future}).  The contents of the data release
are summarized in Table~\ref{table:dr12_contents}, and are described
in detail in the sections that follow for each component survey of the
SDSS-III.  

As described in Section~\ref{sec:boss}, the BOSS spectroscopy is now
complete in two large contiguous regions in the Northern and Southern
Galactic caps.  DR12 represents a $\sim 40\%$ increment over the
previous data release (DR10).  The first public release of APOGEE data
(Section~\ref{sec:apogee})
was in DR10; DR12 represents more than a three-fold increase in the
number of spectra, and six times as many stars with 12 or more visits.
In addition, DR12 includes the first release of data from MARVELS.
MARVELS was in operation for four years (2008--2012); all resulting
data are included in the release.  The MARVELS data (Section~\ref{sec:marvels}) include
$\sim$5,500 unique stars, most of which have 20--40 observations (and thus radial
velocity measurements) per star.  DR11 and DR12 represent different
pipeline processing of the same observed MARVELS data.
The MARVELS fields were selected to have 
$>90$ FGK stars with $V<12$ and $30$ giant stars with $V<11$ in the SDSS
telescope $3^\circ$ diameter field of view.  
A set of pre-selection spectra of these fields to distinguish giants and
dwarfs and thus refine the MARVELS target list was taken by the SDSS
spectrograph in 2008. 
The raw data from these observations were released as part of DR9. In
DR12, we provide the outputs from 
custom reductions of these data.

While \mbox{SDSS-III} formally ended data collection at the end of the night
of 2014 June 30, the annual summer maintenance shutdown at APO
occurred 2014 July 14.  The \mbox{SDSS-III} BOSS and APOGEE targeting
programs were continued during these two weeks 
and are included in the DR12 release. 

In addition, prototype and commissioning data were obtained during
\mbox{SDSS-III} for the \mbox{SDSS-IV} Mapping Nearby Galaxies at APO (MaNGA) project
\citep{Bundy15}, which uses the BOSS spectrographs to measure
spatially resolved spectra across galaxies.  The raw data from
these observations are included in DR12, but reduced data products
(including kinematic and stellar population measurements) will be
released only with the first \mbox{SDSS-IV} data release. 

We also made a single fiber connection from the APOGEE instrument
to the nearby New Mexico State University (NMSU) 1-m telescope at APO for observations when the APOGEE instrument was not
being fed photons from the 2.5-m telescope.   These observations, of a
single star at a time, were
taken to extend the range of the APOGEE-observed stars to
brighter limits, giving improved calibration with existing observations
of these stars \citep[see][for details]{Holtzman15}.
These data and the reductions are included in the standard SDSS-III
APOGEE DR12 products and can be identified by the denoted
source. 

\onecolumngrid
\LongTables
\begin{deluxetable*}{lrrrr}
\tablecolumns{0pt}
\tablecaption{Contents of DR11 and DR12 \label{table:dr12_contents}}
\tablehead{
              &  \multicolumn{2}{c}{DR11} & \multicolumn{2}{c}{DR12} \\
& \colhead{Total} & \colhead{Unique\tablenotemark{a}} & \colhead{Total} & \colhead{Unique\tablenotemark{a}} 
}
\startdata
\cutinhead{All SDSS Imaging and Spectroscopy}
Area Imaged\tablenotemark{b} [deg$^2$]  & & &      31637 &     14555 \\ 
Cataloged Objects\tablenotemark{b} & & & 1231051050 & 469053874 \\  
Total spectra                      & & &    5256940 & \\

Total useful spectra\tablenotemark{p}   & & & 5072804 & 4084671 \\

\cutinhead{MARVELS Spectroscopy (Interferometric)}
Plates\tablenotemark{c}      &  1581 & 241 & 1642  & 278     \\

Spectra\tablenotemark{d}         & 189720 & 3533 & 197040 & 5513 \\

\quad Stars with $\geq$ 16 visits & & 2757 & & 3087 \\

\cutinhead{APOGEE Spectroscopy (NIR)}
Plates                    &    1439 &   547 &    2349 &   817 \\ 
Pointings                 & \nodata &   319 & \nodata &   435 \\
& & & & \\

All Stars\tablenotemark{e}                  & 377812  & 110581  &  618080 & 156593 \\
\quad Stars observed with NMSU 1-m          &         &         &    1196 &    882 \\
Commissioning Stars                         &  27660  &  12140  &   27660 &  12140 \\
Survey Stars\tablenotemark{f}               & 353566  & 101195  &  590420 & 149502 \\ 
\quad Stars with S/N$>100$\tablenotemark{g} & \nodata &  89207  & \nodata & 141320 \\

\quad Stars with $\ge3$ visits              & \nodata &   65454 & \nodata & 120883 \\
\quad Stars with $\ge12$ visits             & \nodata &    3798 & \nodata &   6107 \\
\quad Stellar parameter standards           &    7657 &    1151 &    8307 &   1169 \\
\quad Radial velocity standards             &     202 &      16 &     269 &     17 \\
\quad Telluric line standards               &   46112 &   10741 &   83127 &  17116 \\
\quad Ancillary science program objects     &   20416 &    6974 &   36123 &  12515 \\
\quad Kepler target stars\tablenotemark{h}  &   11756 &    6372 &   15242 &   7953 \\

\cutinhead{BOSS Spectroscopy (Optical)}

Spectroscopic effective area [deg$^2$]              & \nodata & 8647 & \nodata & 9376 \\
Plates\tablenotemark{i}                             &         2085 &         2053 &         2512 &         2438 \\
Spectra\tablenotemark{j}                            &      2074036 &      1912178 &      2497484 &      2269478 \\
All Galaxies                                        &      1281447 &      1186241 &      1480945 &      1372737 \\
\quad CMASS\tablenotemark{k}                        &       825735 &       763498 &       931517 &       862735 \\
\quad LOWZ\tablenotemark{k}                         &       316042 &       294443 &       368335 &       343160 \\
All Quasars                                         &       262331 &       240095 &       350793 &       294512 \\
\quad Main\tablenotemark{l}                         &       216261 &       199061 &       241516 &       220377 \\
\quad Main, $2.15 \le z \le 3.5$\tablenotemark{m}   &       156401 &       143377 &       175244 &       158917 \\
Ancillary spectra                                   &       154860 &       140899 &       308463 &       256178 \\
Stars                                               &       211158 &       190747 &       274811 &       247216 \\
Standard Stars                                      &        41868 &        36246 &        52328 &        42815 \\
Sky                                                 &       195909 &       187644 &       238094 &       223541 \\
Unclassified spectra\tablenotemark{n}               &       132476 &       115419 &       163377 &       140533 \\

\cutinhead{SEGUE-2\tablenotemark{b} Spectroscopy (Optical)}
Spectroscopic effective area [deg$^2$]              & & & \nodata & 1317 \\
Plates & & & & 229 \\
Spectra & & & 155520 & 138099 \\

\cutinhead{All Optical\tablenotemark{o} Spectroscopy from SDSS as of DR12}
\multicolumn{3}{l}{Total spectra}                          & 4355200 & \\
\multicolumn{3}{l}{Total useful spectra\tablenotemark{p}}  & 4266444 & \\

\multicolumn{3}{l}{ \quad  Galaxies}                       & 2401952 & \\
\multicolumn{3}{l}{ \quad  Quasars}                        &  477161 & \\
\multicolumn{3}{l}{ \quad  Stars}                          &  851968 & \\
\multicolumn{3}{l}{ \quad  Sky}                            &  341140 & \\
\multicolumn{3}{l}{ \quad  Unclassified\tablenotemark{n} } &  200490 &
\enddata

\tablenotetext{a}{Removing all duplicates, overlaps, and repeat visits from the ``Total'' column.}
\tablenotetext{b}{These numbers are unchanged since DR8.}
\tablenotetext{c}{
Number of plate observations that were successfully processed through the respective pipelines.  
}
\tablenotetext{d}{Each MARVELS observation of a star generates two spectra.  Unique is number of unique stars.}
\tablenotetext{e}{2,155 stars were observed during both the
  commissioning period and the main survey.  Because commissioning
  and survey spectra are kept separate in the data processing, 
  these objects are counted twice in the Unique column. 
}
\tablenotetext{f}{The statistics in the following indented lines
  include only those observations which met the requirements of being survey quality.}
\tablenotetext{g}{Signal-to-noise ratio per half resolution element
  $>100$, summed over all observations of a given star.}
\tablenotetext{h}{Kepler stars were originally targeted by APOGEE under an
  ancillary program, but eventually became part of the main target selection.}
\tablenotetext{i}{Repeated observations of plates in BOSS are from the 
Reverberation Mapping program (\citealt{Shen15a}, including 30
observations of a single set of targets to study variability),
several other ancillary programs, and several calibration programs.}
\tablenotetext{j}{This count excludes the small fraction ($\sim 0.5\%$) of the observations
  through fibers that are broken or that fell out of their holes after
  plugging.  There were attempted observations of 2,512,000 BOSS spectra.}
\tablenotetext{k}{``CMASS'' and ``LOWZ'' refer to the two galaxy
  target categories used in BOSS \citep{DR9}.  They are both color-selected, 
  with LOWZ galaxies targeted in the redshift range $0.15 < z < 0.4$, 
  and CMASS galaxies in the range $0.4 < z < 0.8$.}
\tablenotetext{l}{This counts only quasars that were targeted by the
  main quasar survey \citep{Ross12},
  and thus does not include those from ancillary programs:
  see Section~\ref{appendix:boss_ancillary}, \citet{Dawson13}, and \citet{Paris14}.}
\tablenotetext{m}{Quasars with redshifts in the range $2.15<z<3.5$
  provide the most signal in the BOSS spectra of the
  Lyman-$\alpha$ forest.}
\tablenotetext{n}{Non-sky spectra for which the automated
redshift/classification pipeline \citep{Bolton12} gave no reliable
classification, as indicated by the \code{ZWARNING} flag.}
\tablenotetext{o}{Includes spectra from SDSS-I/II (DR7;
  \citealt{DR7}).  Although the MARVELS interference spectra are in the
  optical range (5000\AA$<\lambda<$5700\AA), for convenience of
  labeling we here differentiate between the MARVELS data as
  ``interferometric'' and the original SDSS or BOSS spectrograph data
  as ``optical.''} 
\tablenotetext{p}{Spectra on good or marginal plates.}
\end{deluxetable*}

\twocolumngrid

\section{MARVELS}
\label{sec:marvels}

The MARVELS survey (J.~Ge et al.~2015, in preparation) was designed to
obtain a uniform census of radial-velocity-selected planets around
a magnitude-limited sample of F, G, and K main sequence, subgiant, and giant stars.  It aimed to
determine the 
distribution of gas giant planets ($M>0.5$~$M_{\rm Jupiter}$) 
in orbits of periods
$<2$~years and to explore the ``brown dwarf desert'' over the mass range
$13<M<80~M_{\rm Jupiter}$ \citep{Grether06}.  
Measuring these distributions requires a target sample
with well-understood selection and temporal sampling.  
These science goals
translated to observational plans to monitor 8400 stars over 2--4
years with radial velocity accuracies of $10$--$50$~m~s$^{-1}$ for
$9<V<12$~mag for each of 24 epochs per star.  These radial velocity accuracy
predictions were estimated as 2 times the theoretical photon-noise limit.  

The MARVELS instrument~\citep{Ge09}, the W. M. Keck Exoplanet Tracker,
uses an innovative dispersed fixed-delay interferometer
(DFDI) to 
measure stellar radial velocities, by observing the movements of
stellar lines across the fringe pattern created by the interferometer.
The wavelength coverage of the interferometer is
$\rm 5000\AA<\lambda<5700\AA$
and it simultaneously observes 60 science fibers.

MARVELS radial velocities (RVs) are
differential measurements, based on the shift of a star's fringing
spectrum at the current epoch relative to one from the template epoch.
 For more details on the
MARVELS program and DFDI instruments
see \citet{Eisenstein11,Erskine00,Ge02a,Ge02b,Ge09,vanEyken10} and 
J.~Ge et al. (2015, in preparation).

As described in \citet{Eisenstein11}, the original plan was to build
two MARVELS spectrographs so as to 
capture 120 stars per exposure and a total sample of 11,000 stars.
However, due to lack of funding, the second spectrograph was not
built, meaning that the total number of stars observed was about 5500.
We unfortunately encountered significant challenges in calibrating the RV stability of the MARVELS instrument.  
These difficulties led us to end the MARVELS observing as of the summer
shutdown in 2012 July, so as to focus on our data reduction efforts. 
For a detailed accounting and presentation of the observations see
Table~\ref{table:dr12_contents} and Figures~\ref{fig:marvels_numobs}
and \ref{fig:marvels_sky}. 
The typical RMS scatter of the radial velocity measurements in the
data processing we have achieved to date has been 3--5 times greater
than the photon noise limit.  This increased RMS has significantly
limited the ability to discover planets in the MARVELS data. 
However, the distribution of RMS values extends to near
the photon noise limits and has led to cautious optimism 
that further improvements in processing and calibration may 
yield improved sensitivity to giant planets.

The original data processing pipeline was based on software from earlier DFDI prototype instruments \citep[e.g.,][]{Ge06}. 
 This pipeline used the full 2-D phase information but the resulting
 radial velocity measurements were limited by systematic
 instrumental variations to an RMS of 100--200~m~s$^{-1}$.  
 As described in detail below, the two radial velocity estimates from
 this pipeline are presented 
 in DR11 as the ``cross-correlation function'' (CCF) and
 ``differential fixed-delay interferometry'' (DFDI) reductions, the
 latter explicitly incorporating the phase information from the
 interferometric fringes.  
 These reductions revealed instrumental calibration variations that
 required a redesign of the analysis approach.  

A subsequent reworked processing pipeline only
analyzes the collapsed one-dimensional (1-D) spectrum, without using the
fringing information, but determines the
calibration of the spectrograph dispersion on a more frequent basis
(N.~Thomas et al.~2015, in preparation). 
 The results from this pipeline are presented in DR12 as the ``University of Florida One Dimensional'' (UF1D) reductions.

\subsection{Scope and status}

\begin{figure}
\plotone{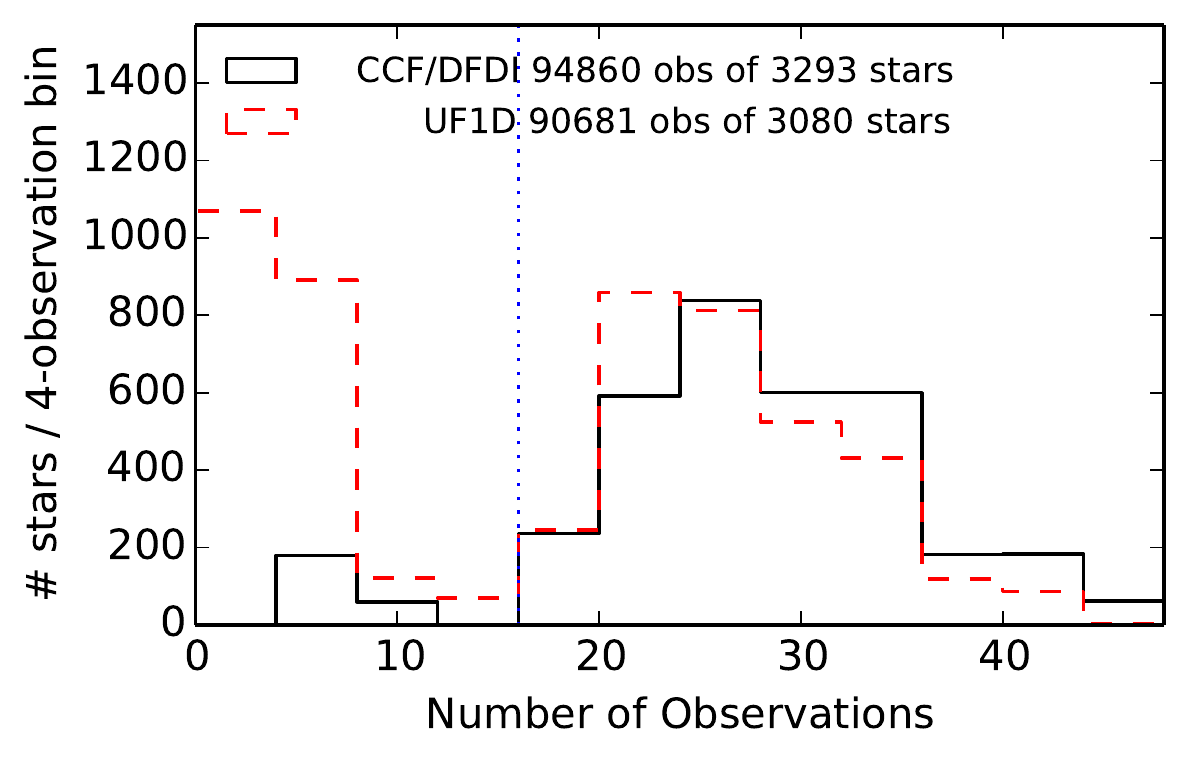}
\caption{Distribution of the number of observations made of each
  MARVELS star that was processed by the CCF+DFDI (black solid) and the UF1D (red dashed) pipelines and met the respective quality cuts.
}
\label{fig:marvels_numobs}
\end{figure}

MARVELS data collection began in 2008 October and ended in 2012 July.
The majority of MARVELS stars were observed 20--40 times
(Figure~\ref{fig:marvels_numobs}), with a typical exposure time of
50--60~min.  These exposure times were designed to reach a
signal-to-noise ratio (SNR)
sufficient to allow per-epoch RV precisions of tens of m~s$^{-1}$ on stars of
$7.6<V<12$~mag.  The total number of observations was designed to
enable the determination of orbital parameters of companions with
periods between one day  
and two years without the need for follow-up
RV measurements using additional telescopes.  However, the problems in
radial velocity calibration, the shortened MARVELS observing period,
and the fact that the second MARVELS spectrograph was never built
meant that this ideal was not met for all targets. 
The observing was
split into two 2-year campaigns: Years 1+2: 2008 October -- 2010
December; and Years 3+4: 2011 January -- 2012 July.  For any particular
star, the time baseline between the first and last observation was
thus typically 1.5--2 years.

During its four years of operation MARVELS obtained 1565 observations
of 95 fields collecting multi-epoch data for 5700 stars, with
observations of 60 stars per target field.

While we provide all raw data and
intermediate data products in this release, the CCF and DFDI results are
limited to the 3533 stars with more than 10 RV measurements.  The UF1D
analysis results include 5513 stars from the 92 fields that pass the basic quality requirements of the pipeline.
Restricting to stars with $\geq16$ observed epochs, which might be considered a 
reasonable threshold for searching for companions in the MARVELS data,
yields 3293 stars in DR11 and 3233 stars in DR12 (a small number
because of somewhat tighter quality constraints). 

\subsection{A Brief Guide to MARVELS Data}

Each spectrographic plate has two sets of 60 fiber holes, corresponding to two
 different fields to be observed in sequence.  
Both sets of fibers were plugged at the same time.  In between
observations of the two fields, the ``gang'' connector that
 joins the fibers from the cartridges to the long fibers that run 
to the MARVELS instruments was switched between the two
 sets of fibers. 

A MARVELS exposure is the result of light from each of 60 fibers
passing through a two-beam interferometer with one slanted mirror
and then dispersed in wavelength before being recorded on a $\rm 4k
\times 4k$ CCD. 
Thus each MARVELS image contains 120 individual spectra as the
beam-splitter produces two interference patterns for each star,
one from each beam.  
The RVs for each star can then be calculated from a comparison of the
fringing spectrum observations at different epochs.  

In this data release we provide the two-dimensional (2-D) raw images,
the 2-D slices of extracted spectra, the 1-D collapsed spectra,
and the calculated stellar velocities and associated 
observational metadata for each spectrum of each star and field.

\subsection{Target selection}

Target selection for MARVELS is described in full in 
\citet{Paegert15}.
 We here summarize the key aspects
of the MARVELS target selection in each two-year phase of the survey. 

MARVELS aimed to have a target sample in the range of $8<V<12$  
with a balance of 90\% dwarf and subgiant stars with $T_{\rm eff} < 6250$~K, 
and $\sim$10\% giant stars with $4300<T_{\rm eff}<5100$~K (spectral
types K2--G5).  
In the first two years of
MARVELS, target selection was based on short ``pre-selection''
observations 
obtained with the SDSS spectrographs during the first year of
\mbox{SDSS-III} to determine stellar surface temperatures and surface
gravities. 
 Because these observations used much shorter exposure times than
standard SDSS observations, they were not automatically processed with the standard SDSS
pipeline.
Instead, the SDSS pipeline was used with some custom modifications  
to provide stellar spectra suitable for processing with the SEGUE Spectroscopic Processing Pipeline \citep[SSPP;][]{Lee08}.
 The raw data for these spectra were released as part of DR9. 
In DR12 we release these custom spectroscopic images, extracted spectra, 
and derived SSPP parameters as flat files, but due to their specialized 
and non-standard nature these have not been loaded into the CAS. 

Unfortunately, the derived $\log{g}$ values --- needed
to discriminate giants from dwarfs --- from these
moderate-resolution spectra ($R\sim2000$) were not reliable and the
first two years of MARVELS targets resulted in a 35\% giant fraction
instead of the goal of 10\%.  

We thus employed 
a new method for giant-dwarf selection in Years 3+4.
For this second phase of the MARVELS survey,
temperature estimates were 
derived based on $V-K$ and $J-K$ colors following the 
infrared flux method of \citet{Casagrande10},
and giants were rejected based on a requirement of
a minimum reduced proper motion \citep{CollierCameron07} based on 
the measured 2MASS $J$-band proper motion
together with the $J$-band magnitude and $J-H$ color.

From 2011 January onward all MARVELS observations were carried out
simultaneously with APOGEE, using plug plates drilled with holes for
both sets of targets.  The spectroscopic cartridges were adapted to allow
connection of both the APOGEE and MARVELS fibers to the long fibers
that run to the stabilized rooms that house the respective instruments.  
 This joint observation mode yielded
significant overall observational efficiencies, but imposed the
restriction that both surveys observe the same fields with the same
cadence.  This shifted the MARVELS target fields much farther south
than originally planned as
APOGEE pursued observations toward the center of the Milky Way. 

\begin{figure*}
\plotone{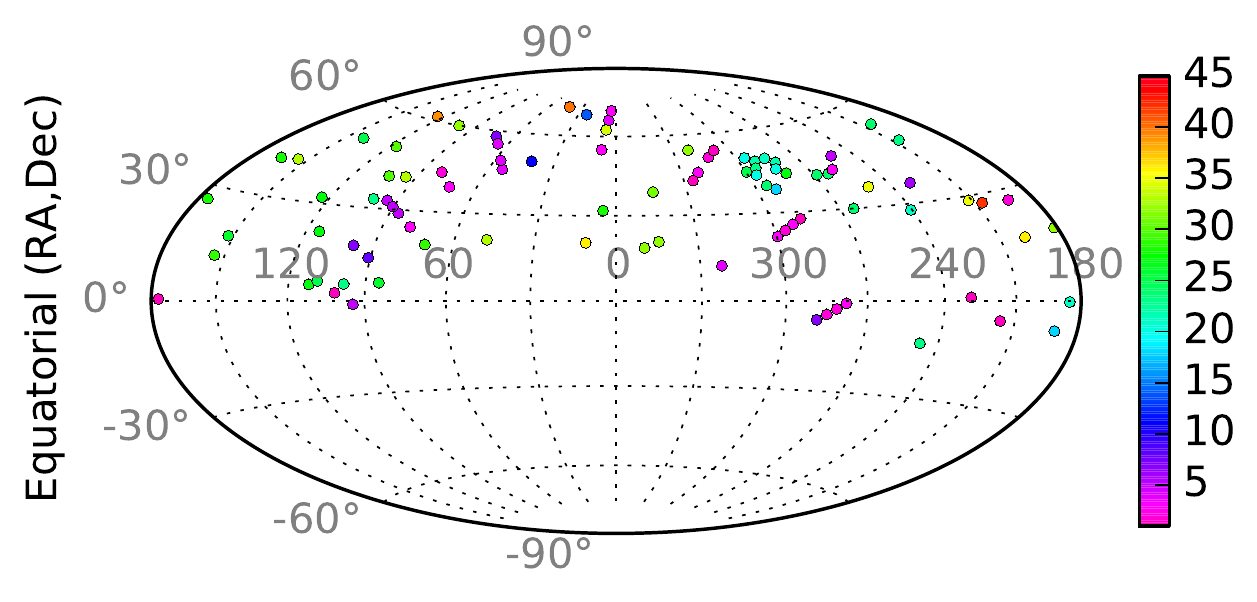}
\caption{MARVELS sky coverage in equatorial coordinates.  Each plate
  is plotted with a color-coding giving the number of epochs the plate
  was observed. 
}
\label{fig:marvels_sky}
\end{figure*}

The sky distribution of all observed MARVELS fields is shown
in Figure~\ref{fig:marvels_sky}.  

\subsection{MARVELS Data Analysis}

\begin{figure*}
\includegraphics[width=4in]{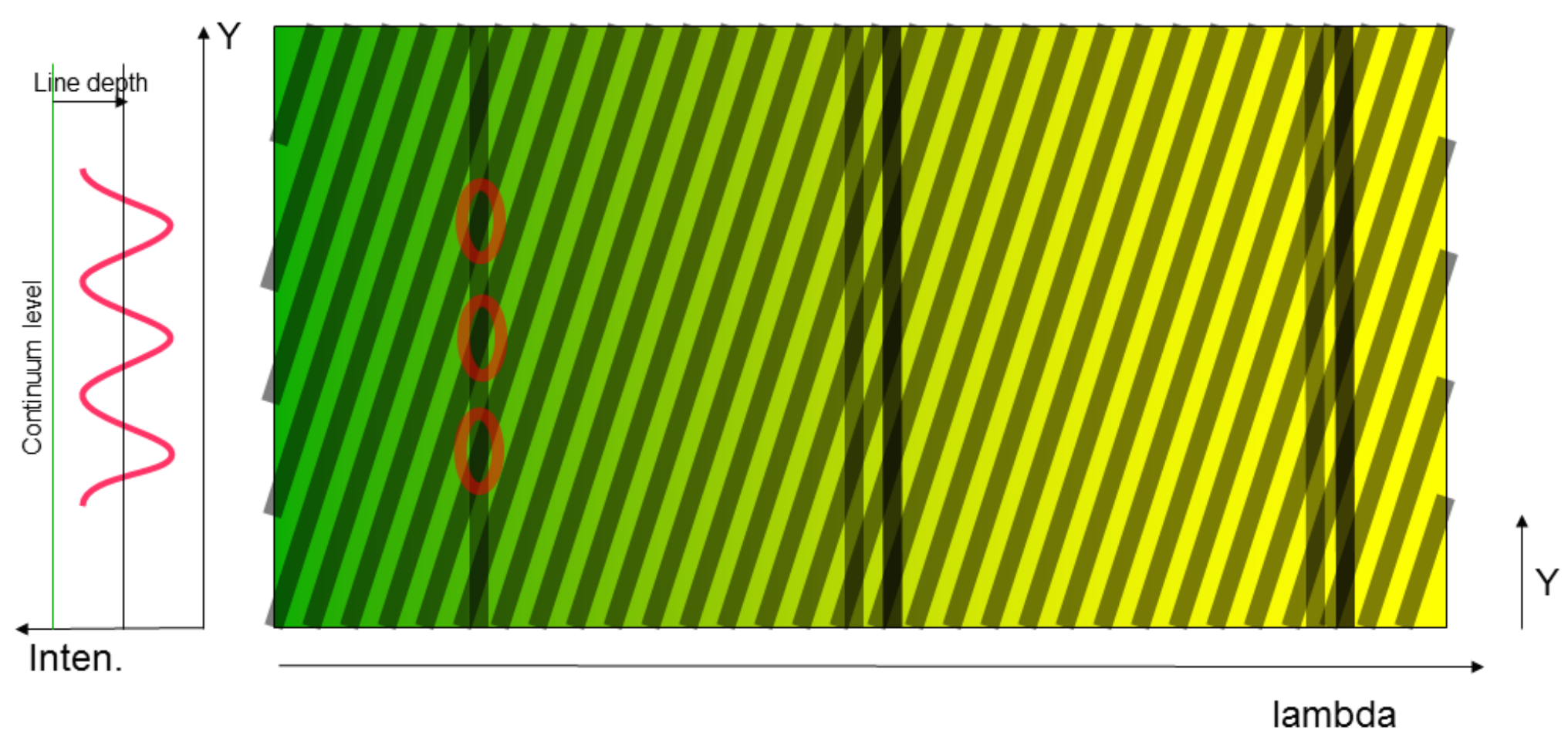}
\includegraphics[width=2.7in]{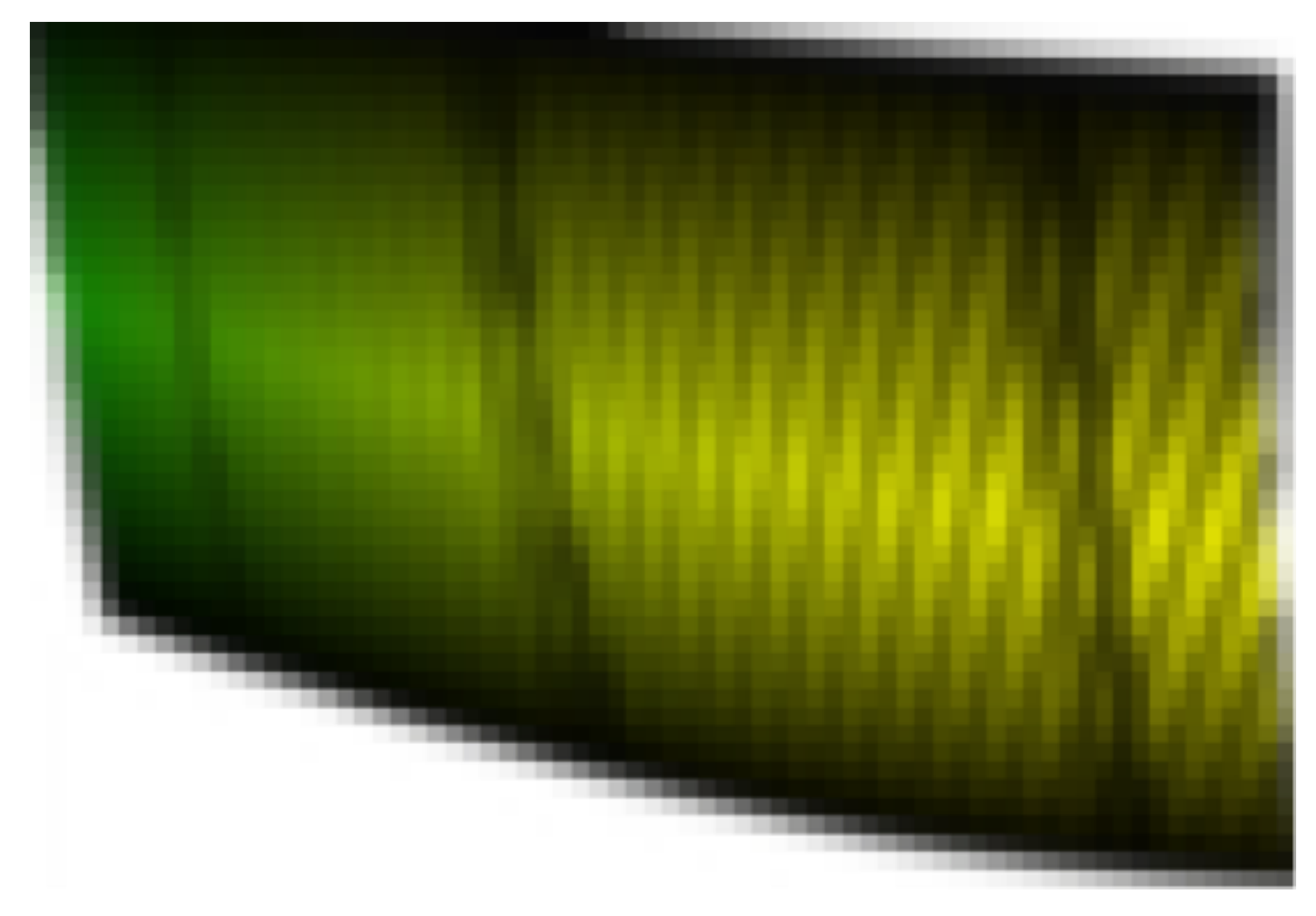}
\caption{
(left) Conceptual illustration of a portion of the spectrum of one star 
from MARVELS dispersed fixed-delay interferometry. 
For simplicity, we show only five absorption lines in this sample wavelength region; the full MARVELS wavelength range features thousands of absorption lines -- most of these are blended at the MARVELS dispersion of $R\sim11,000$.
The diagonal pattern of constructive and destructive
interference is not sharp as in this simple diagram, but rather varies
sinusoidally with $y$. The phase of the
best-fitting sinusoid to each column of the data determines the
corresponding wavelength shift, given the slope of the interference
comb. 
(right) Illustration of 
some of the real-world effects of variable projection of spectra onto the focal plane,
spectrograph alignment, point spread function, and the variable
slope of the interference comb. 
Note the additional blending in each set of closely-separated absorption lines.
There are 
120 of these spectra (each roughly 4096 pixels by 34 pixels) per
MARVELS exposure. 
}
\label{fig:marvels_interference}
\end{figure*}

The MARVELS instrument is designed to be sensitive to wavelength
shifts (and thus radial-velocity changes) in stellar spectra.  It
splits each input stellar spectrum into two beams, and then
projects a slanted interference pattern of the recombined beams 
through a spectrograph (see Figure~\ref{fig:marvels_interference}).  

The dispersed slanted interference pattern effectively magnifies the resolution of a
moderate-resolution spectrograph ($R\sim11,000$) by
translating wavelength shifts in the dispersion (``$x$'') direction to much larger
shifts in the ``$y$'' position.
This slope is $\sim5$ pixel~pixel$^{-1}$ for MARVELS.  
 The design goal of the MARVELS analysis
is to measure the shift of the interferometric sinusoid in the $y$
direction to determine the wavelength offset due to a radial velocity change. 

The key challenges in
the processing of MARVELS data are the calibration of the wavelength
solution on the detector, identification and extraction of each
spectrum, and the measurement of the slant of the interferometric comb
and of the resulting interference pattern of the absorption-line features.

Our approach to analyzing the MARVELS data 
will be described in detail in
N.~Thomas et al. (2015, in preparation),
which specifically describes the UF1D pipeline.
The CCF+DFDI and UF1D pipelines follow many of the same steps,
but differ in choices of calibration reference sources
and complexity of model for instrumental variations.
We here outline the important differences in the CCF+DFDI and UF1D processing.

\subsubsection{Extraction of Spectra from the 2-D Images}
A key part of spectroscopic processing is determining the ``trace'',
i.e., where the light from a given fiber target falls on the CCD.  In
an idealized instrument, the trace would lie horizontally along the
CCD (constant $y$), and the light at a given wavelength would be
distributed perpendicular to the trace (constant $x$), In practice,
this is not true, and we correct for these two according through a
``trace correction'' and ``deslant correction''.

The CCF+DFDI pipeline uses available Tungsten lamp continuum exposures
with a diffuser 
to determine the trace of the spectrum on the CCD, and Thorium-Argon 
arc spectra to determine the deslant correction.
The UF1D pipeline uses the Tungsten lamp exposures taken through an
iodine cell to determine the trace, and the absorption lines in the
observed stellar spectra to determine the deslant correction.
The pipelines extract and correct 2-D arrays for each spectrum
based on their respective trace and deslant corrections.

\subsubsection{Compression to One-Dimensional Spectra}

The CCF+DFDI pipeline takes the 2-D rectified spectrum
and fits a sinusoid to the interference pattern along the $y$ (slit) direction.
The spectrum is then collapsed along $y$, and the resulting 1-D spectrum plus
sinusoidal fit parameters are stored. The combination of the collapsed
spectrum and the sinusoidal fits 
is denoted a ``whirl'' in the provided CCF+DFDI data products.

The UF1D pipeline focuses on improvements to the instrumental calibration
without adding complications from the details of the phase
extraction.  It simply collapses the 2-D rectified spectra 
along the $y$ direction to create 1-D spectra, removing 
the information contained in the fringes. 
The UF1D pipeline was implemented as a step toward a new pipeline
still in development that will include 
the more detailed calibration model used in the UF1D pipeline (see
below) 
and will also make use of the phase information from the 2-D spectra.

\subsubsection{Characterizing the Instrumental Wavelength Drift}
Determining the instrumental wavelength drift over time is critical in
deriving reliable radial-velocity measurements.  The instrumental
drift is measured from calibration lamp exposures taken before and
after each science frame.  The calibration exposures are from a
Tungsten lamp shining through a temperature-stabilized Iodine gas cell
(TIO).  This extracted spectrum is compared to that of the calibration
lamp exposures taken on either side of the reference epoch chosen as
the baseline for that star.

For the CCF+DFDI pipeline, the shift for each star was determined by
comparing the extracted TIO spectrum to a single reference lamp
spectrum taken on MJD 55165 (2009 November 29),  
and the measured radial velocity for the star in question was
corrected by the resulting offset.  This correction attempts to express all
changes in the instrument by a single parameter per fiber.
The large variance in the resulting radial velocities has shown that
this approach does
not fully capture the complex nature of the calibration changes across
the detector.  

In an effort to capture the fact that the velocity offset may be a
function of wavelength, the UF1D pipeline calculates
a separate shift value for each 100-pixel chunk of each spectrum,
corresponding to $\sim 17$\AA. 
The reference TIO pair for each field is chosen to be the one that
brackets the observation with the highest stellar flux
observations. These instrumental shift values are then used as corrections
to each chunk of the spectrum before the stellar radial
velocity shifts are determined.

\subsubsection{Measuring the Stellar Radial Velocity Shifts}
In CCF+DFDI, the stellar radial velocity is measured by comparing the
extracted stellar spectrum from a given stellar exposure to 
the spectrum at the template epoch.
The template epoch is selected as the highest SNR
observation available for the selected star.  
 We first calculate the barycentric correction (due to the orbit of
 the Earth around the Sun) as part of the
 comparison with the template epoch, and then use cross-correlation to
 measure the radial velocity offset of the 1-D spectrum.  
This raw stellar radial velocity shift is 
corrected for the instrumental drift determination from the previous
step and labeled as the CCF measurement. 
The fringe shifts as a function of wavelength are then used to refine
these velocity offsets to generate the final DFDI measurements.  
These two successive calculations are reported in separate tables in
DR11 with CCF and DFDI suffixes in the name of the
respective tables. 

In principle, the DFDI radial velocities should be more
precise.  However, given the challenges in measuring stable radial
velocities from the processing, we find it useful to compare the
results with (DFDI) and without (CCF) the fringe corrections. 

In UF1D, the pixel shift of each stellar spectrum with respect to that
from the template date is determined for the same 100-pixel chunk based on
a least-squares solution that minimizes the difference in values in each pixel, and then corrected for the calibration drifts
measured from the TIO measurements.  The resulting calibrated shifts
are converted into a radial-velocity measurement by using a wavelength
solution from each 100-pixel chunk to covert from pixel shift to
wavelength shift to velocity shift.  The outlier-rejected mean
velocity shift across all 100-pixel chunks is then taken as the
velocity shift for that spectrum for that epoch.

These radial velocity shifts are then corrected for the barycentric
motion of each observation.  Because the radial velocity measurements
are all relative, the zero point of the radial velocities is
meaningless, so the mean of all measurements for a given star is set
to zero.  

Because of the two-beam nature of the DFDI instrument, each star
observation results in two spectra.  These computations are done
separately for each of these two spectra.  
The published data tables present RVs separately for each beam.
To estimate the RV for the star on a given epoch one would 
in principle simply average the
radial velocities from the two measurements.
Because of the noticeable number of outliers in individual beam measurements, 
the use of an outlier rejection scheme is recommended.

\begin{figure*}
\plottwo{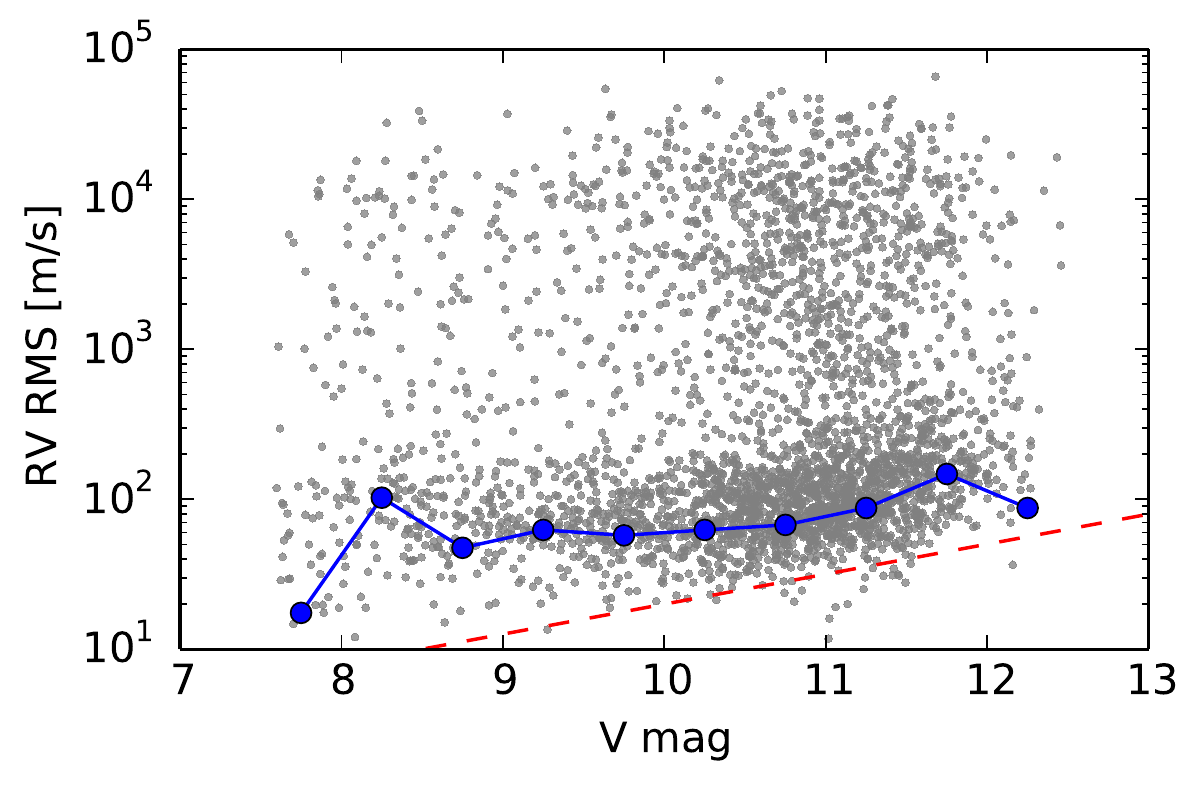}{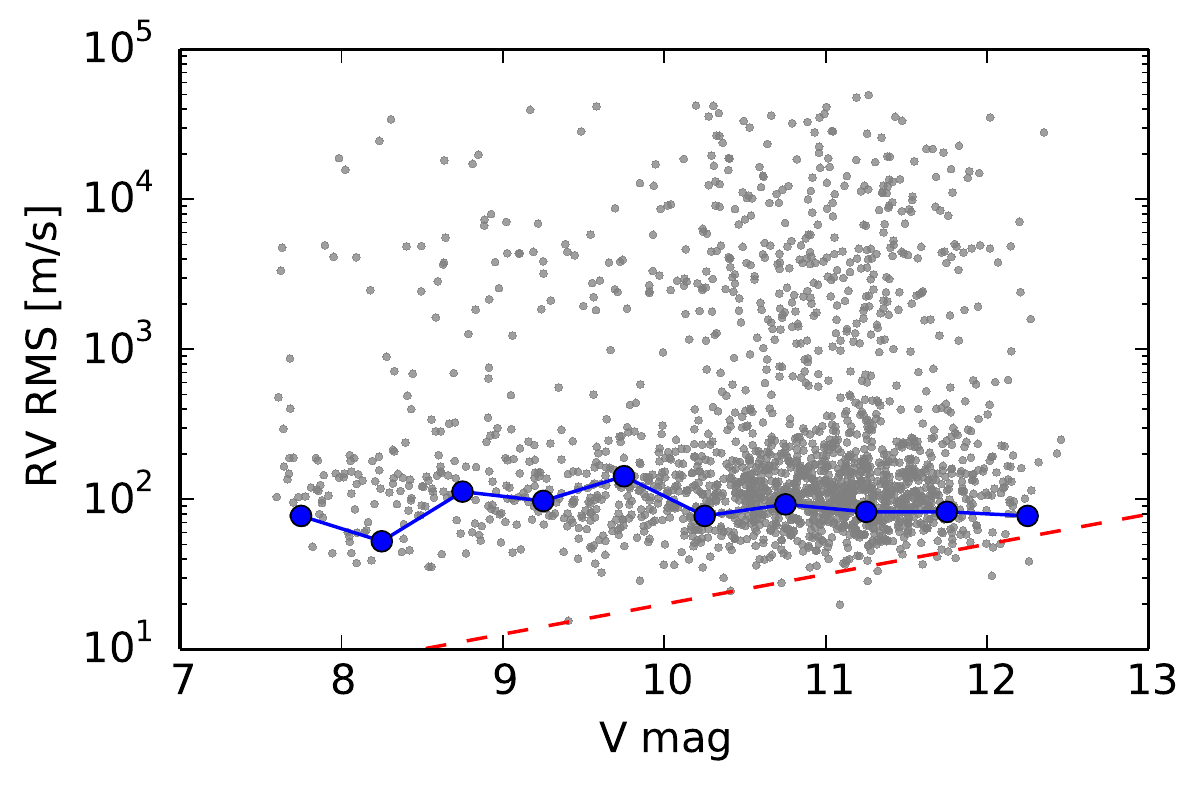}
\caption{Distribution of RMS of radial velocity measurements of
 MARVELS stars (grey points) for the DFDI (left), and UF1D (right) analyses, 
 as a function of apparent magnitude.  
 The mode of the RMS in each 0.5~mag bin (blue circles and line)
 highlights the significant number of stars with RMS near 50--100 m~s$^{-1}$.
However, a comparison with the theoretical photon limit (red dashed line) 
illustrates that the bulk of the RMS values are many
  times higher than the limit. 
Despite this, there are stars whose radial velocity
  repeatability approaches the theoretical limit, suggesting that the
  large scatter for many of the observations is due to calibration
  problems, which might be improved with further development of the
  pipeline.  }
\label{fig:marvels_mag_rms}
\label{fig:marvels_mag_rms_all}
\end{figure*}

\begin{figure*}
\plottwo{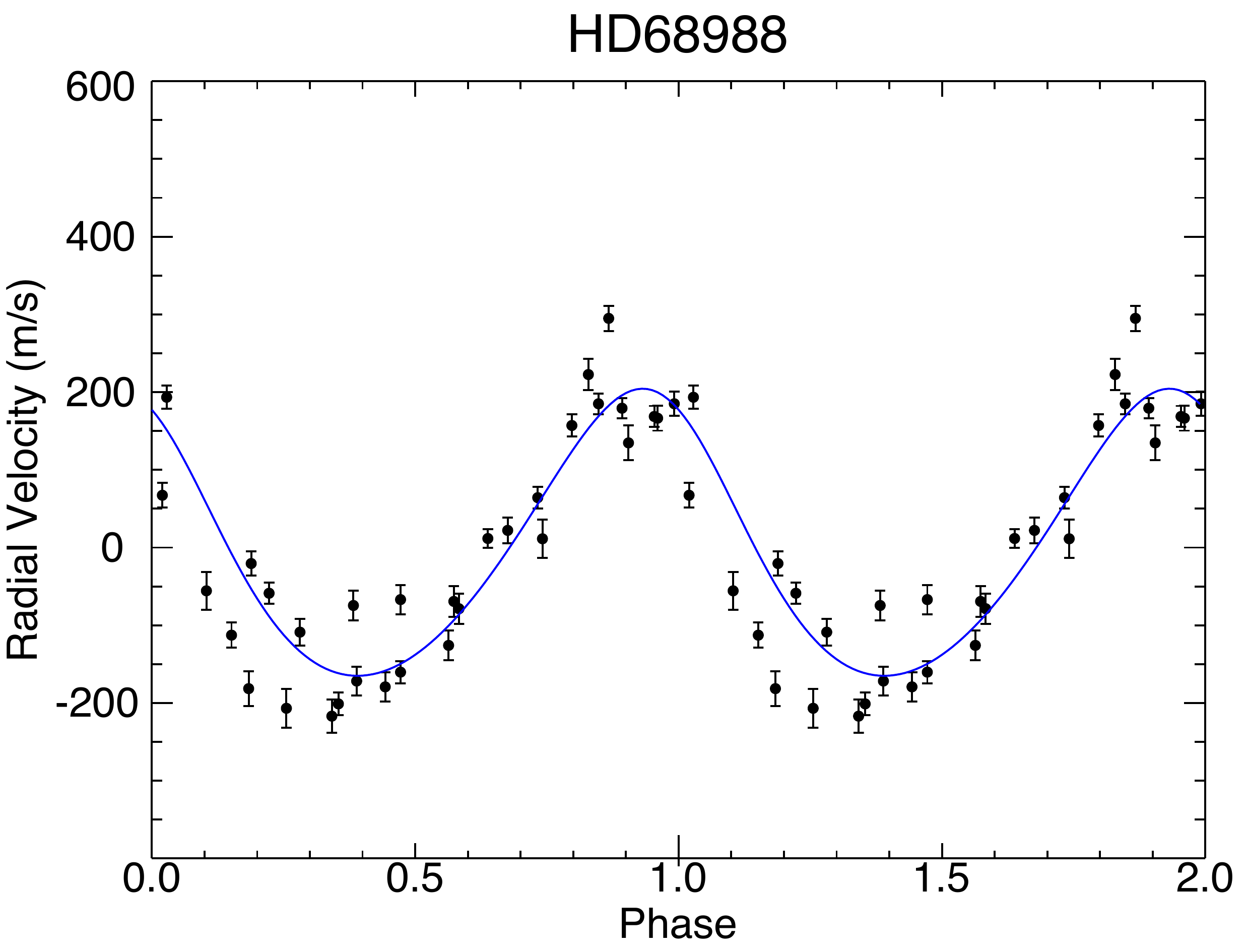}{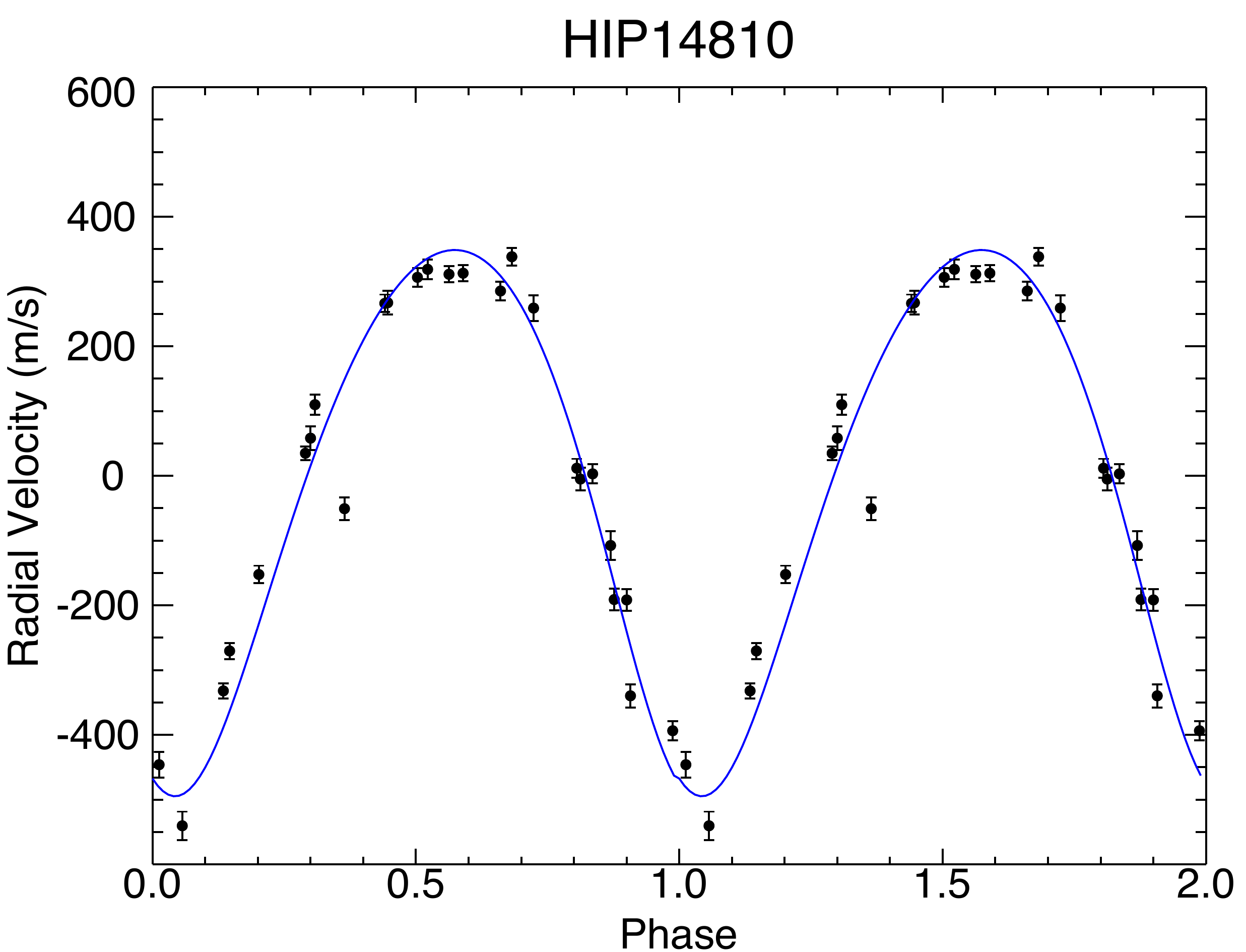}
\caption{MARVELS observations of the radial velocities of the stars 
(left)  HD68988 compared to the exoplanet model of \citet{Butler06}; and 
(right) HIP-14810 compared to the model of \citet{Wright09}.
The phased data are shown over two periods for ease of visualization.
}
\label{fig:marvels_known_systems}
\end{figure*}

\subsection{Current Status and Remaining Challenges}

As Figure~\ref{fig:marvels_mag_rms} shows, the current data processing results in
stellar radial velocity variations of 50 m~s$^{-1}$ or
larger even at high SNR, a value several times
greater than that expected from photon statistics.  This is mostly due to
systematic uncalibrated wavelength shifts on timescales longer than a
month; repeat observations of stars within the same lunation show much
smaller radial velocity variations.  However, the figures show that
{\em some} stars show RMS radial velocity variations which approach
the photon noise limit, suggesting that with proper calibration, the
overall scatter should drop significantly.  
One possibility currently under investigation is that these stars represent specific fibers
that are more stable, while the beams from other stars 
experienced greater hardware variation across repeated pluggings 
and fiber connections.  Work continues on improving the analysis of
the MARVELS data 
and our understanding of the long-term systematic effects.

Despite these challenges, the MARVELS DR11 reductions have been used
to study low mass and sub-stellar companions
\citep{Wisniewski12,Fleming12,Ma13}, brown dwarfs in the ``desert''
\citep{Lee11}, and exotic orbital systems \citep{Mack13}. 
Figure~\ref{fig:marvels_known_systems} shows MARVELS RV measurements
of two stars with known exoplanets, showing that MARVELS data are
in good 
agreement with existing orbital models for these systems.

However, in general the MARVELS data and analysis to date have not
achieved the survey requirements for radial velocity  necessary to discover and
characterize a fiducial 0.5-$M_{\rm Jupiter}$ planet in a 100-day orbit.
Figure~\ref{fig:marvels_mag_rms} 
shows the achieved radial velocity RMS for the current pipelines as a
function of stellar magnitude. 
The upper band of objects with RMS from 1--10 km~s$^{-1}$ is
predominantly true astrophysical variation from binary star systems.
The distribution of objects with RMS values in the range of 100 m~s$^{-1}$ 
is bounded near the photon limit, but the bulk lies several times
above these limits.

\section{BOSS}
\label{sec:boss}

\subsection{Scope and Summary}
\label{sec:BOSS_scope}

The BOSS main survey of galaxies and quasars over two large contiguous
regions of sky in the Northern and Southern Galactic Caps was completed in 
Spring 2014.  The majority of the galaxies were uniformly targeted for
large-scale structure studies in a sample
focused on relatively low redshifts (``LOWZ'', with $z < 0.4$) and a
sample with $0.4 < z < 0.7$ designed to give a sample approximately
volume-limited in stellar mass
(``CMASS''; B.~Reid et al.~2015, in preparation).  The total footprint
is about 10,400 deg$^2$ 
(Figure~\ref{fig:boss_coverage}); the value of 9376 deg$^2$ in
Table~\ref{table:dr12_contents} excludes masked regions due to bright stars
and data that do not meet our survey requirements.

The main BOSS survey was completed in 2014 February.  The additional dark
time available through the 2014 summer 
shutdown was devoted to a portfolio of additional science programs
designed to maximize the science return while taking advantage
of the unique abilities of the SDSS system.
Two of the largest such programs were
a variability study of 849 quasars, designed to measure time delays
between continuum and emission line variations (``Reverberation
Mapping''; \citealt{Shen15a}),  and an early start on the planned
cosmological studies with \mbox{SDSS-IV} (the Sloan Extended QUasar, ELG and LRG
Survey, hereafter ``SEQUELS'', where ``ELG'' stands for ``Emission
Line Galaxy'' and ``LRG'' stands for ``Luminous Red Galaxy''), together with
an exploratory set of plates to investigate the requirements for
studies of high-redshift ELGs and other aspects of SDSS-IV. These and
other BOSS ancillary programs executed since the DR10 release are described in
Appendix~\ref{appendix:boss_ancillary}.   

\subsection{Highlights from BOSS DR11}
\label{sec:BOSS_DR11}

The DR11 and DR12 releases of BOSS data constitute increments of 35\% and
47\% in the number of spectra over DR10, respectively, processed using
very similar pipelines.  These increases were significant enough to
warrant a new set of BOSS cosmological analyses for each of these
releases.  These key papers were one of the motivations for tagging a
DR11 data set for later public release along with DR12.  The cosmology
analyses based on DR11 data 
include studies of isotropic galaxy clustering \citep{Guo15},
 anisotropic galaxy clustering \citep{Song14,Samushia14,Sanchez14,Gil-Marin14a,Gil-Marin14b,Reid14,Beutler14a},
 galaxy clustering in the LOWZ sample (\citealt{Tojeiro14}), 
 the baryon oscillations (BAO) in
 the clustering of the Lyman-$\alpha$ forest of distant quasars \citep{Bautista14,Delubac15}, 
 the first detection of BAO in the cross-correlation between the Lyman-$\alpha$ forest and the quasars \citep{Font-Ribera14},
 an updated upper bound to the sum of neutrino masses \citep{Beutler14b},
 a summary BAO galaxy clustering analysis paper \citep{Anderson14},
 and a joint cosmology analysis paper incorporating 
 all of the BOSS cosmology constraints as well as those from Type Ia supernovae 
 and anisotropies in the cosmic microwave background \citep{Aubourg14}. 
 The BOSS team plans a similar set of papers based
on the full DR12 analyses.  

\subsection{Data Reduction Changes for DR12}

The pipeline software for reduction of BOSS spectroscopic data was 
largely unchanged between DR10 and DR11.  The classification and
redshift-measurement aspects of this software are described in 
\citet{Bolton12}.

There were, however, some significant improvements to 
the spectrophotometric flux calibration routine for DR12.  These improvements were
made to mitigate low-level imprinting of (primarily) Balmer-series
features from standard-star spectra onto science target spectra.  This
imprinting was first documented in \citet{Busca13} in observed-frame
stacks of quasar continuum spectra.  Although this effect is generally
undetectable in any single-spectrum analysis, it has a small but
non-negligible effect on the analysis of the Lyman-$\alpha$ forest
across many thousands of quasar spectra.  The change implemented for
DR12 consists of a simple masking and linear interpolation of the
flux-calibration vectors over the observed-frame wavelength ranges
shown in Table~\ref{table:BalmerFix}.  A more flexible flux-calibration
vector model is retained at other wavelengths to accommodate real
small-scale features in the spectrograph throughput.  This more
flexible model was necessary for the original SDSS spectrographs due
to time variation in the dichroic filters, although it is likely
unnecessary for the improved optical coatings on those surfaces in
BOSS \citep[see][]{Smee13}.

In addition, we updated the pixel-response flats used to pre-process
the spectrograph frames, improved the bias-subtraction code to
catch and correct electronic artifacts that appear in a small number
of frames, and updated the CCD bad-pixel and bad-column masks to
reduce the incidence of corrupted but previously unflagged spectra.
These changes reduce the number of corrupted spectra, and more accurately 
flag those that remain.  

Table~\ref{table:BOSS_changes} gives the full history of
significant changes to the BOSS spectrograph detectors and the
calibration software to process its data since the BOSS survey began.
See also Table 2 of \citet{DR9} for additional changes to the
hardware.  

\begin{deluxetable}{cl}
\tablecaption{Wavelength Ranges Masked During BOSS Spectrophotometric Calibration\label{table:BalmerFix}}
\tablehead{
\colhead{Line} & \colhead{Wavelength Range} \\
\colhead{}     & \multicolumn{1}{c}{\AA} }
\startdata
H$\varepsilon$ & 3888.07\,$\pm$\,25 \\
$[\mathrm{Ne} \, \mbox{\textsc{iii}}]$ & 3969.07\,$\pm$\,30 \\
H$\delta$ & 4100.70\,$\pm$\,35 \\
H$\gamma$ & 4339.36\,$\pm$\,35 \\
H$\beta$  & 4860.09\,$\pm$\,35 \enddata
\tablecomments{Observed-frame vacuum wavelength ranges that were masked and linearly interpolated during determination of
spectrophotometric calibration vectors.}
\end{deluxetable}

\begin{deluxetable*}{lll}
\tablecaption{Significant changes to the BOSS
  spectrographs and the data reduction pipeline\label{table:BOSS_changes}}
\tablehead{
\colhead{Date}&MJD&\colhead{Comments}
}
\startdata
2010 April 14 &55301&R2 Detector changed following electrical failure\\
&&R2 pixel flat, bad pixel mask on all four cameras updated\\
2010 August &55410& Bad pixel mask updated on all four cameras\\
&&Pixel flat updated on R1 and R2\\
2011 August &55775& R1 detector changed following electrical failure\\
&&R1 pixel flat, bad pixel mask on all four cameras updated\\
2011 October 16&55851&R1 bad pixel mask updated\\
2012 August &56141& Bad pixel mask updated on all four cameras\\
&&Pixel flat updated on R1 and R2\\
2013 August &56506& Pixel flat updated on R1 and R2\\
2013 December 23 &56650& R2 detector had an electrical failure, but recovered\\
&&R2 bad pixel mask and pixel mask updated\\
2014 February 10 &56699& R1 pixel flat updated
\enddata
\tablecomments{There are two BOSS spectrographs, each with a red and
  blue camera.  Thus R2 refers to the red camera on the second
  spectrograph, which accepts light from fibers 501--1000. The August
  dates in the table above refer to the summer shutdowns.}
\end{deluxetable*}

As in previous BOSS data releases, a unique tag of the
\texttt{idlspec2d} spectroscopic pipeline software is associated with
each unique sample of publicly released data.\footnote{SDSS data processing software is publicly available at \url{http://www.sdss.org/dr12/software/products/}}  
 Three tagged
reductions of three separate samples are being released at the time
of DR12.  One (\texttt{v5\_6\_5}) is the ``DR11'' version that
defines a homogeneous sample of BOSS data taken through Summer 2013;
this is the version used in the cosmological analyses described in
Section~\ref{sec:BOSS_DR11} above.
A second label (\texttt{v5\_7\_0}) defines the main DR12 BOSS cosmological
survey at its point of completion.  A third tag (\texttt{v5\_7\_2})
is associated with the several extra observing programs undertaken
with the BOSS spectrographs in Spring 2014 following the completion of
the main BOSS survey program (Section~\ref{sec:BOSS_scope},
Appendix~\ref{appendix:boss_ancillary}).  These data-release software
versions 
are summarized in Table~\ref{table:PipelineVersions}. 

Many of the pipeline changes for the ancillary programs involved
bookkeeping and special cases for plates drilled with either fewer or
more flux calibration stars.  In addition the SEQUELS plates targeted
ELGs at high redshift, so the
upper redshift limit of the galaxy template fitting \citep{Bolton12} was extended
from $z=1$ to $z=2$.  Thus DR12 includes several
thousand SDSS galaxy spectra with tabulated redshifts above $z=1$.

\begin{deluxetable*}{lll}
\tablecaption{Spectroscopic pipeline versions associated with each BOSS data release. 
 \label{table:PipelineVersions}}
\tablehead{
\colhead{Data Release} & \colhead{Code Version} & \colhead{Comments} }
\startdata
DR8  &      \nodata      & No BOSS spectroscopic data \\
DR9  & \texttt{5\_4\_45} & First BOSS spectroscopic data release \\
DR10 & \texttt{5\_5\_12} & Also includes data first released in DR9 \\
DR11 & \texttt{5\_6\_5}  & Also includes data first released in DR10 \\
DR12 & \texttt{5\_7\_0}  & Main BOSS sample, also includes data first released in DR11 \\
DR12 & \texttt{5\_7\_2}  & Extra BOSS programs, non-overlapping with \texttt{v5\_7\_0} \enddata
\end{deluxetable*}

\begin{figure*}
\plottwo{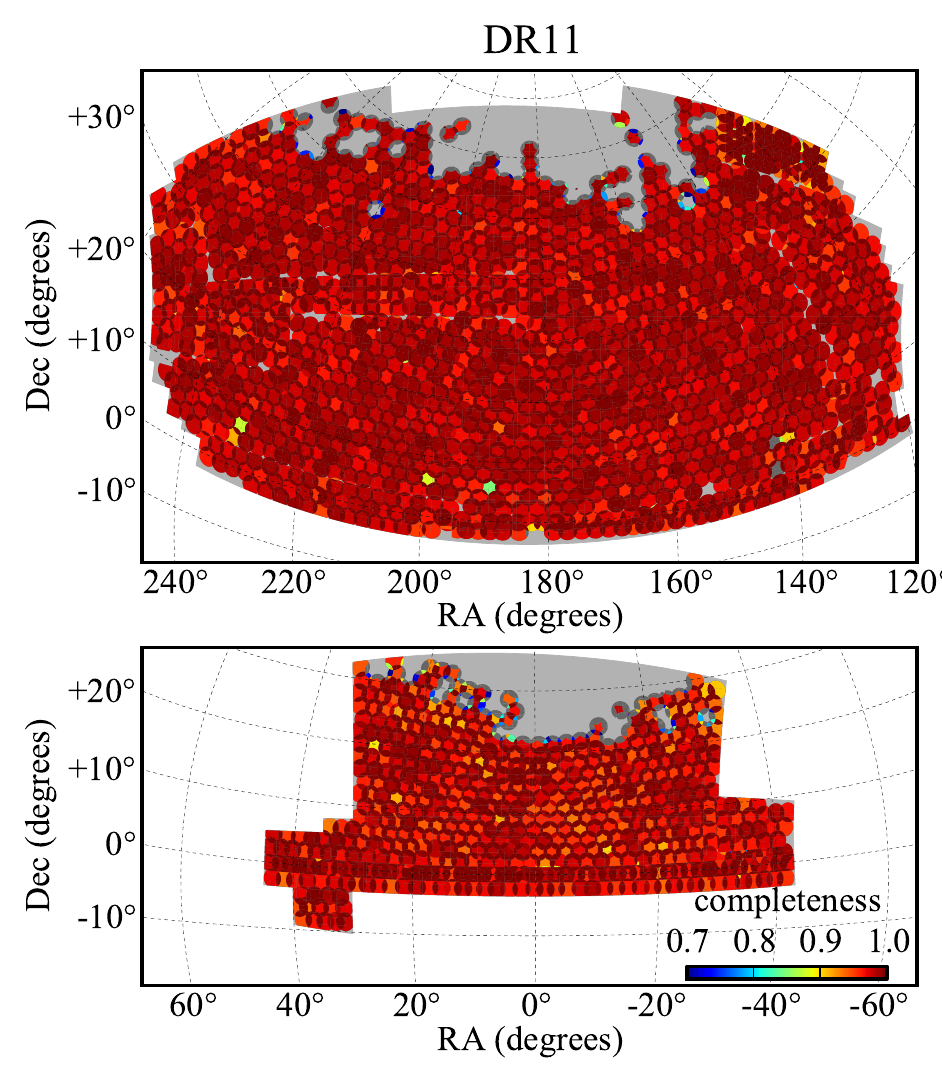}{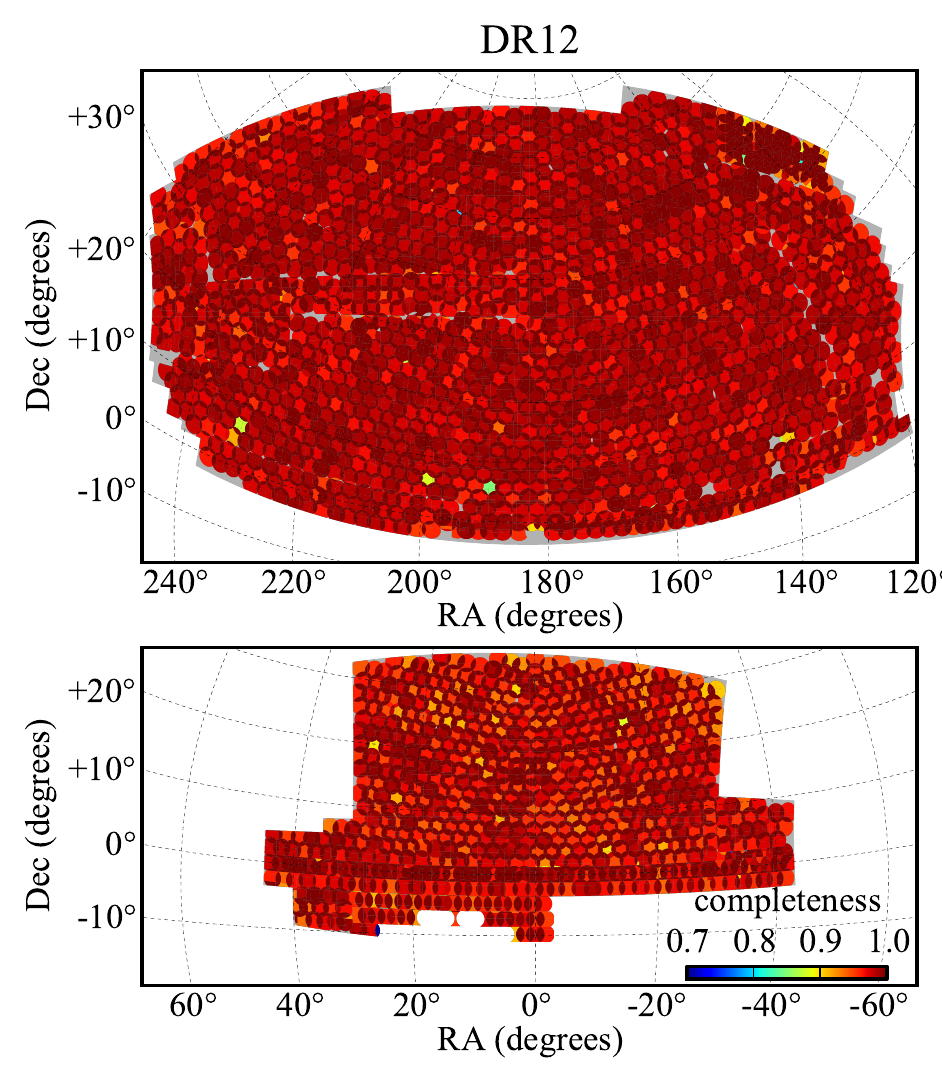}
\caption{BOSS DR11 (left) and DR12 (right) spectroscopic sky coverage in the Northern Galactic
Cap (top) and Southern Galactic Cap (bottom).  The grey region (visible most clearly in the DR11 map) was the coverage goal for the final survey.
The DR12 coverage map shows that we exceeded our original goals with a
final total of 10,400~deg$^2$. The color coding indicates the fraction of
CMASS galaxy targets that receive a fiber.  The average completeness
is 94\% due to the limitation that no two fibers can be placed
closer than $62$\arcsec\ on a given plate. 
\label{fig:boss_coverage}}
\end{figure*}

\section{APOGEE}
\label{sec:apogee}

In this paper, we release both DR11 and DR12 versions of the APOGEE
outputs, with considerably more stars (see
Table~\ref{table:dr12_contents}) in the latter.  The APOGEE release is
described in detail in \citet{Holtzman15}.  The DR11 parameters
and abundances use the same version of the APOGEE
Stellar Parameters and Chemical Abundances Pipeline 
(ASPCAP; A. E. Garc\'ia P\'erez et al.~2015, in preparation)
as in DR10.  The DR12 version of ASPCAP is a major upgrade, in which
abundances are determined for 15 individual elements.  In addition,  
the DR12 ASPCAP code incorporated a number of technical improvements: 
multiple searches to avoid local minima in parameter space, new model
atmospheres with 
updated solar reference abundances and non-solar Carbon- and
$\alpha$-element-to-Iron abundance ratios \citep{Meszaros12},
the use of a Gauss-Hermite function instead of a Gaussian to represent
the instrumental point-spread function, and upgrades to the atomic and
molecular line lists. 
These improvements do not change the derived fundamental stellar
parameters systematically, but do improve their accuracy.

\begin{figure*}
\plotone{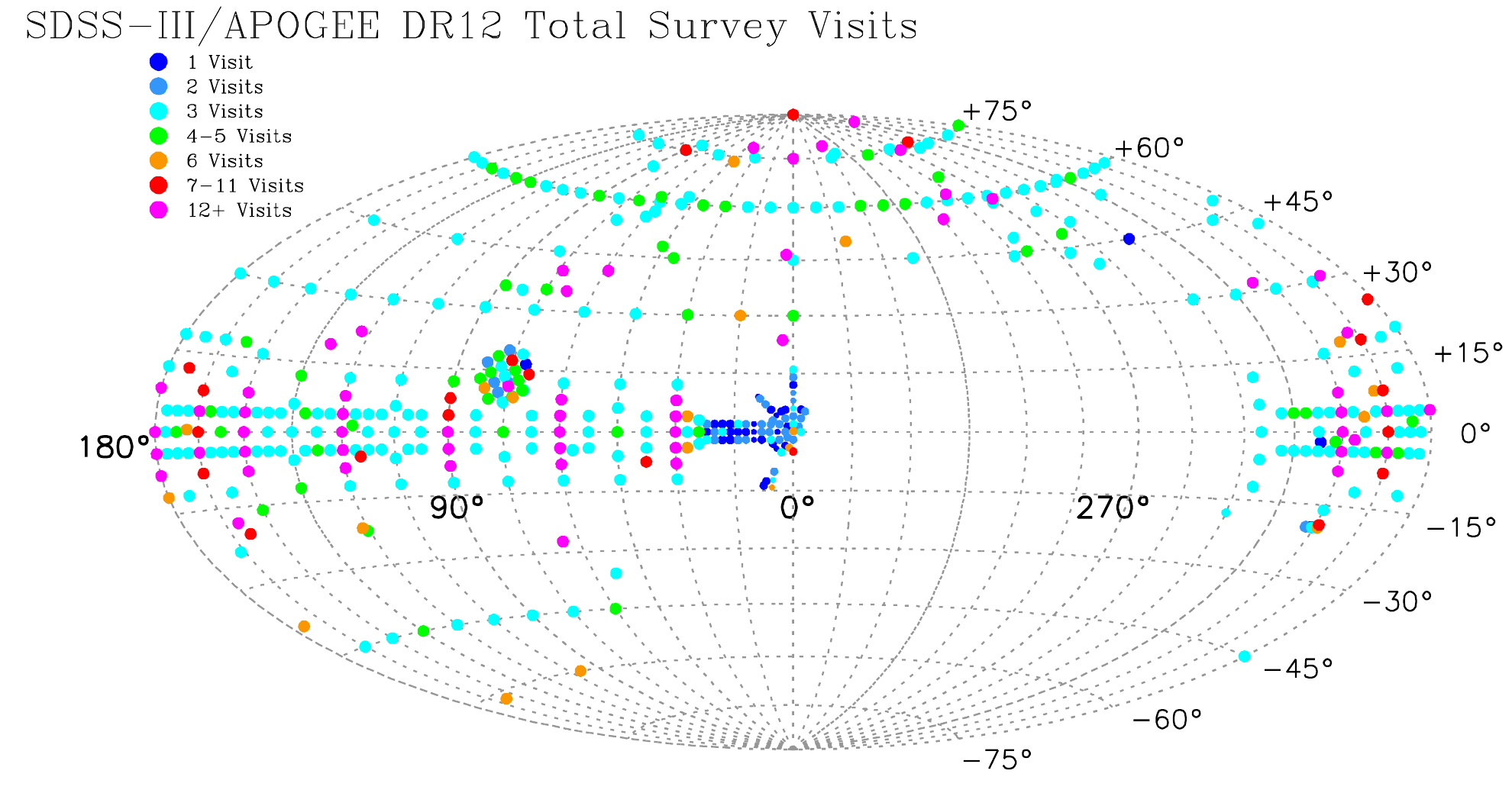}
\caption{Sky coverage of APOGEE DR12 observations in Galactic
  coordinates.  The number of visits to each field is denoted by the
  color coding from 1 visit (blue) through 12 or more visits
  (magenta).} 
\label{fig:APOGEE_DR12_sky_coverage}
\end{figure*}

\begin{figure*}
\plottwo{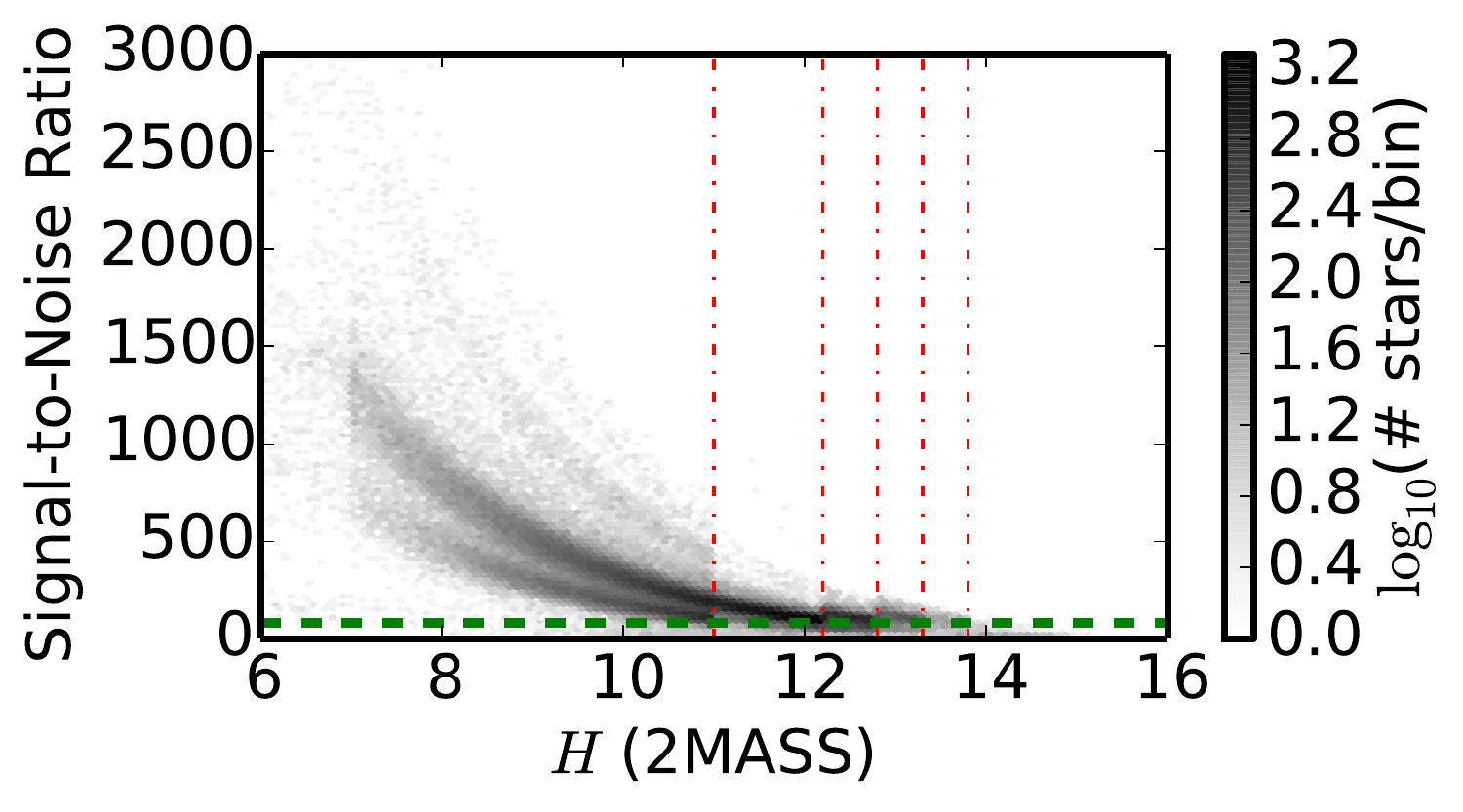}{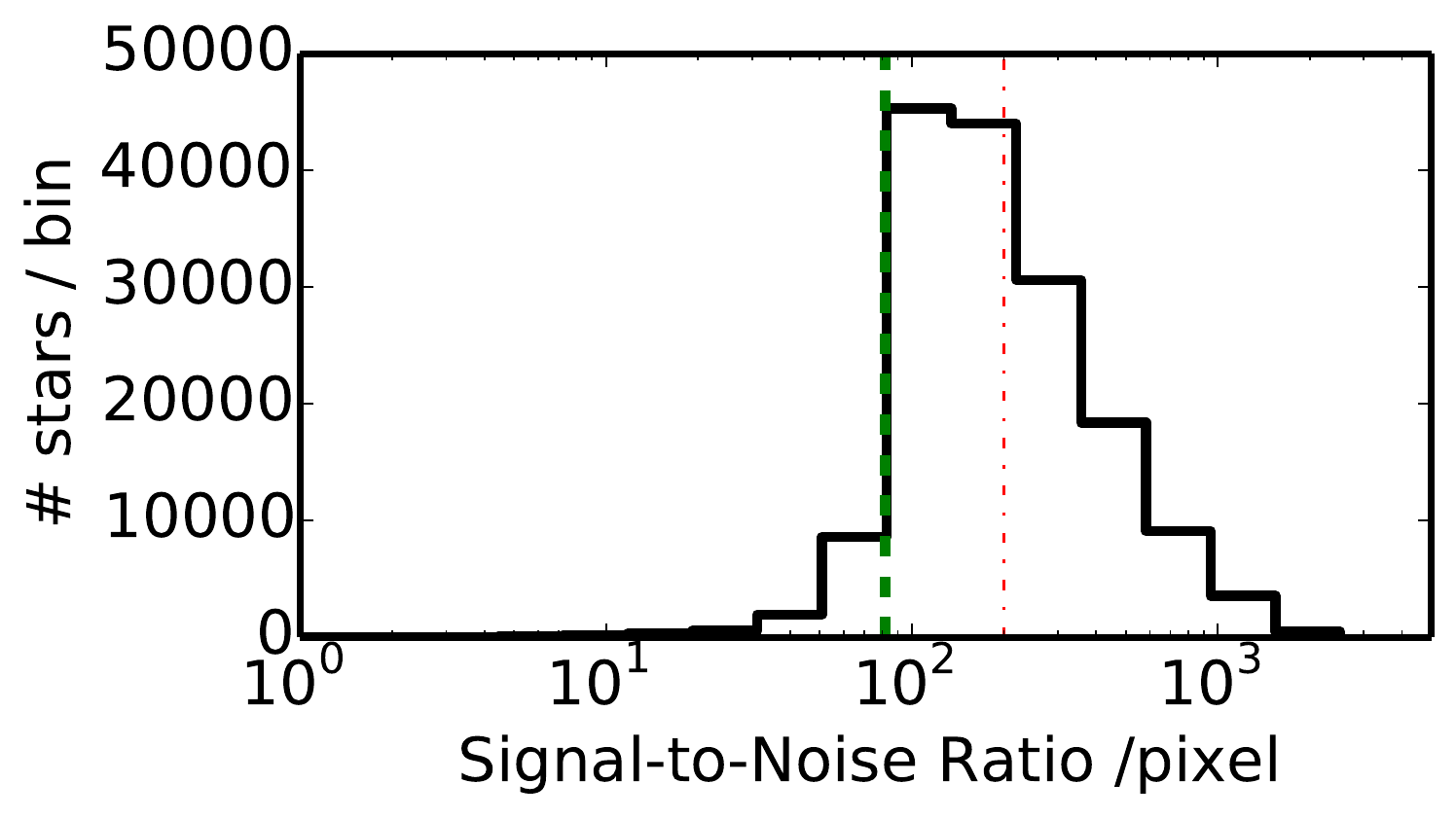}
\caption{
Distribution of SNR of APOGEE stars in DR12.
With 1.5~pixels per effective half-resolution element, the science requirements goal of SNR~$\geq100$/half-resolution element is achieved with SNR~$\geq82$/pixel (dashed green line).
(left) 2-D histogram of SNR vs. 2MASS $H$ magnitude.
The red dash-dot lines denote the magnitude limits for the different bins of target brightness.  The number of planned visits to APOGEE main targets was (1, 3, 6, 12, 24) visits for $H<(11.0, 12.2, 12.8, 13.3, 13.8)$~mag.
(right) 1-D histogram of SNR.
The systematic floor in the effective SNR is $\sim200$ (red dash-dot line).
}
\label{fig:APOGEE_DR12_snr}
\end{figure*}

\begin{figure*}
\plottwo{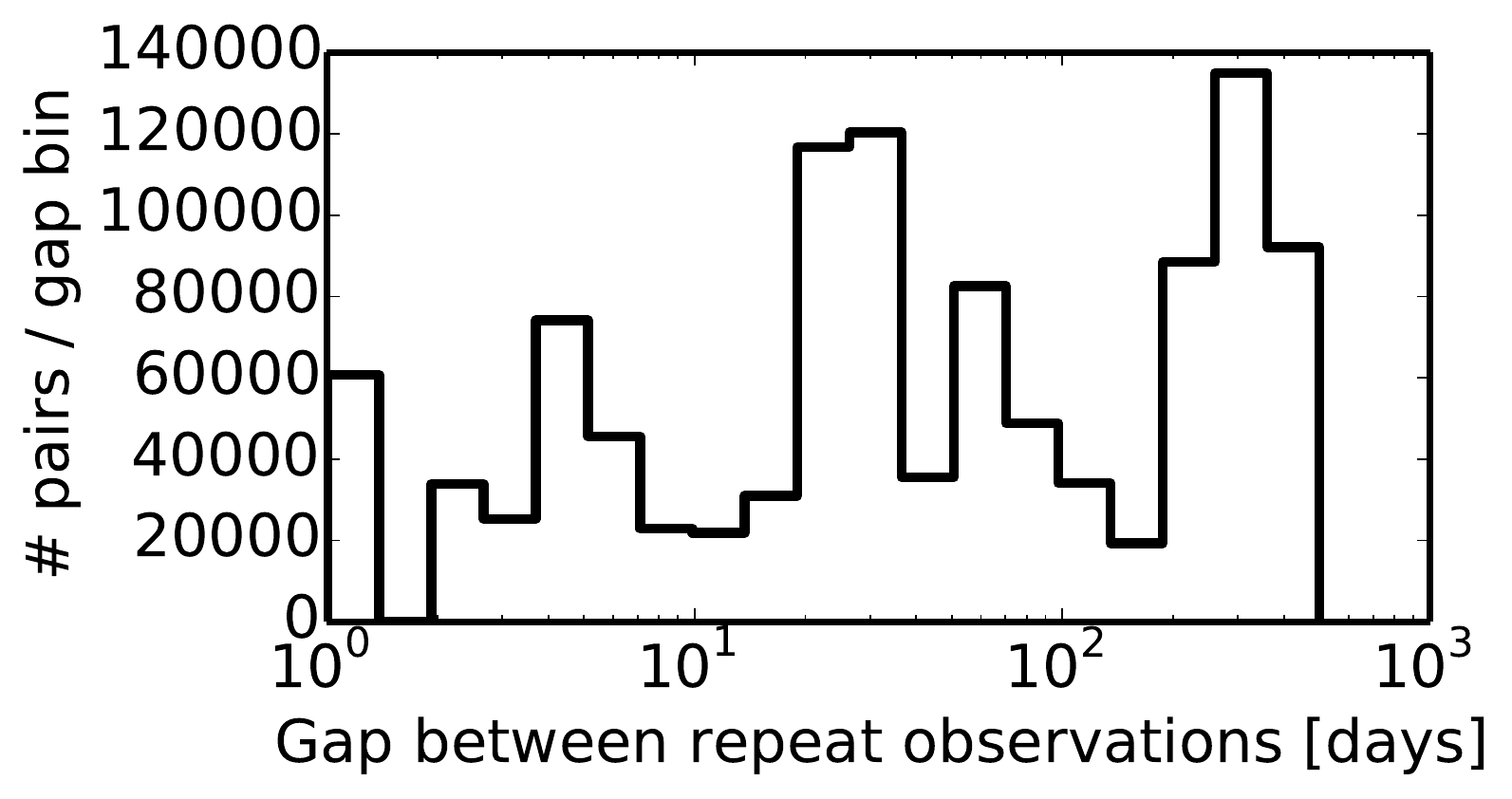}{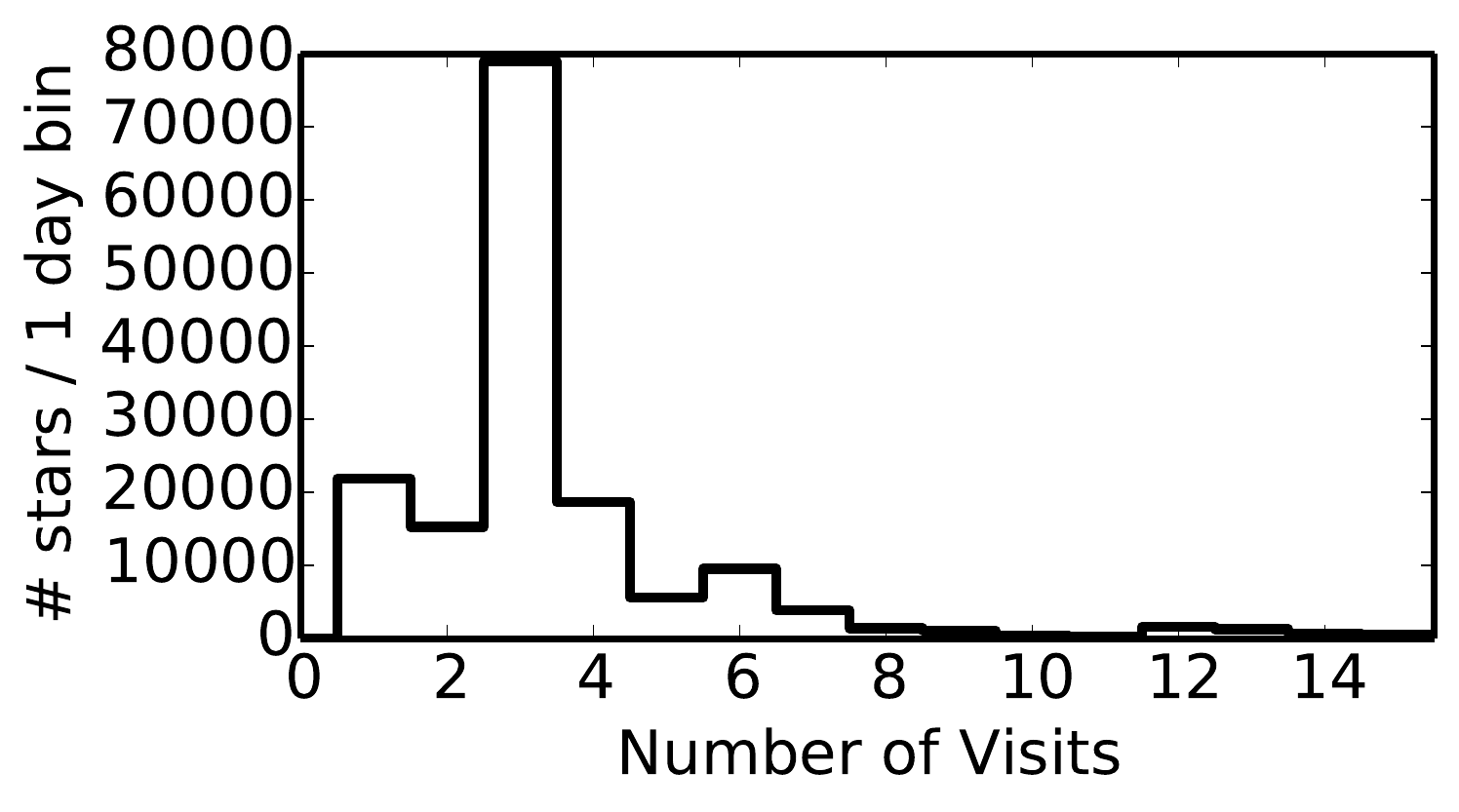}
\caption{
(left) Distribution of time intervals between observations of a given
  APOGEE target in DR12. 
(right) Distribution of number of visits for individual APOGEE targets in DR12.
}
\label{fig:APOGEE_DR12_cadence}
\end{figure*}

\begin{figure*}
\plotone{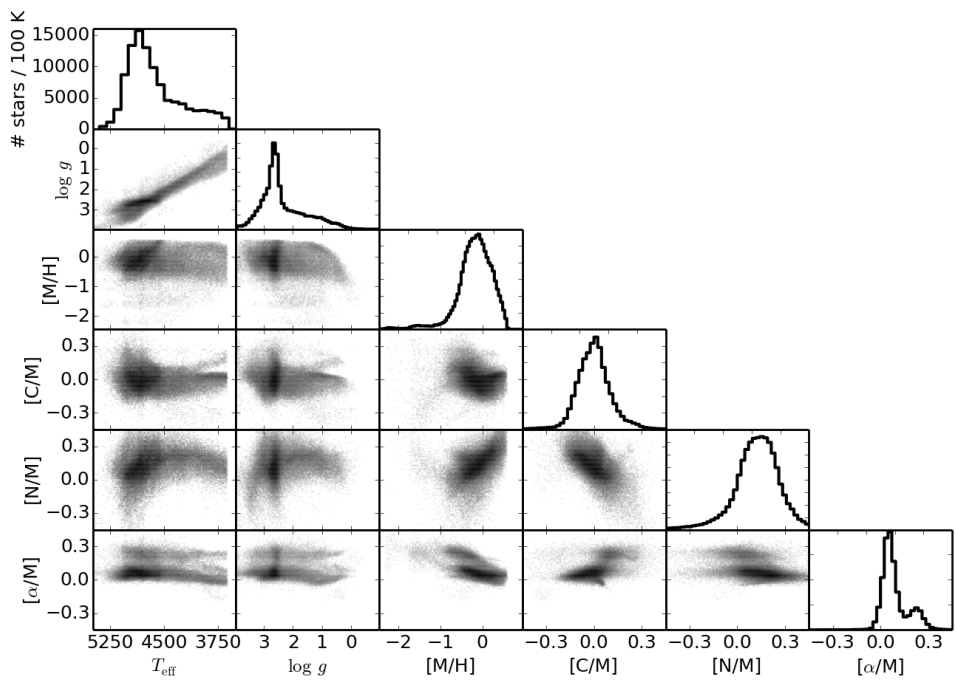}
\caption{
Key stellar parameters ($T_{\rm eff}$,
$\log{g}$) and key metallicity indicators ([M/H], [C/M], [N/M],
[$\alpha$/M]) for stars with APOGEE observations in DR12.  
These
distributions are strongly affected by the selection of stars targeted
for APOGEE spectroscopy.  
The grey scale is logarithmic in number of stars.
}
\label{fig:APOGEE_DR12_logg_Teff_metals_cm_nm_alpham_nn2}
\end{figure*}
  
\begin{figure}
\plotone{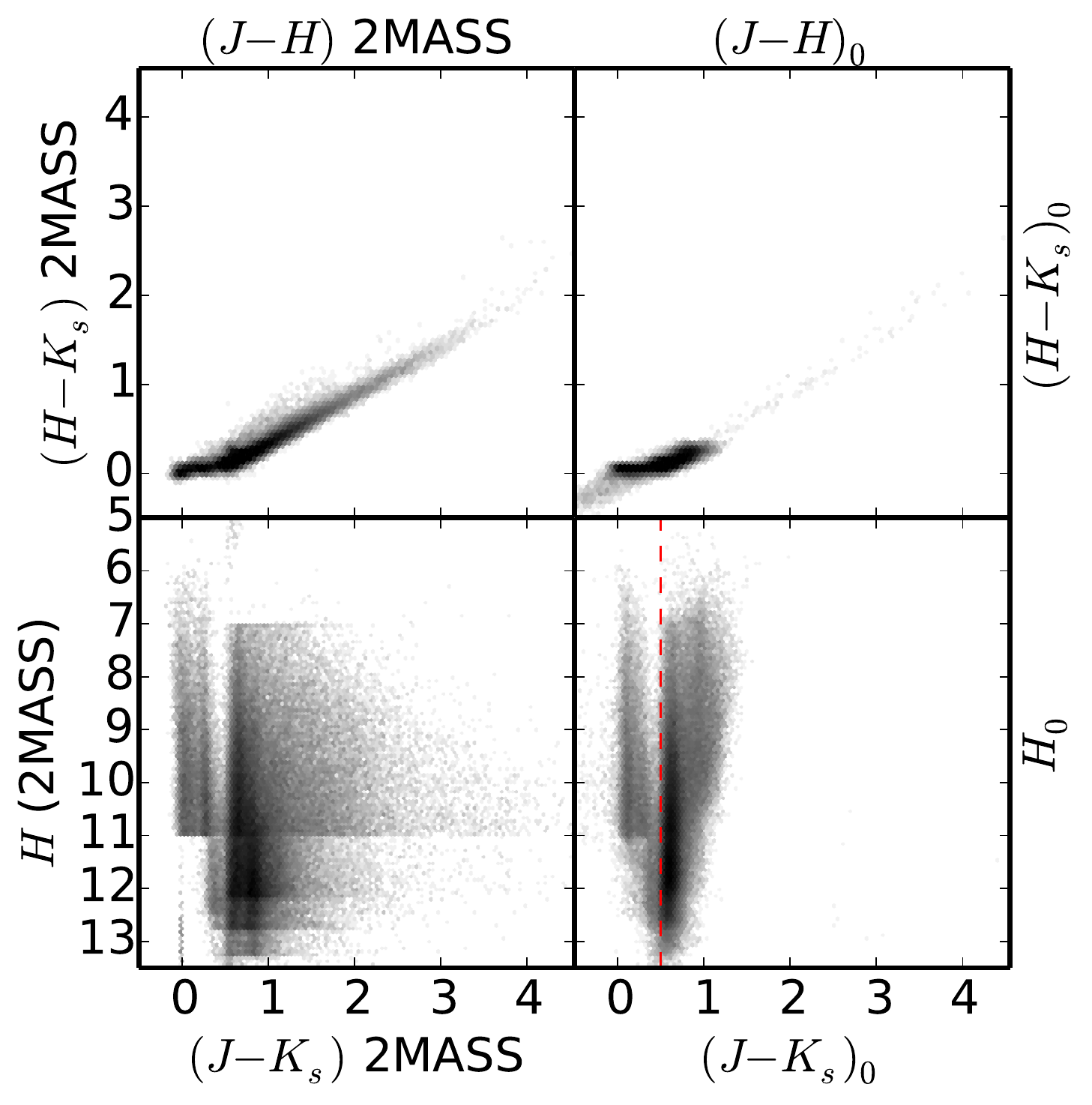}
\caption{
Near-infrared colors and $H$ magnitudes of APOGEE targets
as observed (left panels) and corrected for Galactic dust extinction (right panels).
The vertical dashed line in the lower-right panel at $(J-K_s)_0 = 0.5$~mag 
indicates the selection cutoff for the main APOGEE red giant sample.
Objects bluer than this line are from observations of telluric
calibration stars,  
commissioning data, or ancillary program targets.  
The grey scale is logarithmic in number of stars.
}
\label{fig:APOGEE_DR12_star_color_dereddened_4panel}
\end{figure}

\subsection{Scope and Summary}

The APOGEE DR11 data include twice as many stars and spectra as DR10 (53,000
more stars and 200,000 more spectra), analyzed with the same pipeline.
The APOGEE DR11 data have been used in several papers,
 including a determination of distances to and chemical abundances of red-clump stars \citep{Bovy14,Nidever14},
 mapping of the Galactic interstellar medium using diffuse interstellar bands measured along the line of sight to APOGEE stars \citep{Zasowski15},
and an identification of new Be stars and their $H$-band line profiles \citep{Chojnowski15}. 

APOGEE DR12 represents a further year of data and thus includes another
46,000 stars and 240,000 spectra over DR11.  It also uses the updated
analysis pipeline described above.  
The sky coverage of the final APOGEE DR12, covering
the bulge, disk, and halo of our Galaxy 
is shown in Figure~\ref{fig:APOGEE_DR12_sky_coverage}.  
The additional observations of stars that already appeared in DR10
improve the SNR of these stars and 
also provide opportunities for studies of radial velocity and other variations 
in the observed stellar spectra.  Figure~\ref{fig:APOGEE_DR12_snr} demonstrates
that we achieved our goal of SNR$>100$ per half-resolution element
for the APOGEE sample.
Figure~\ref{fig:APOGEE_DR12_cadence} shows
the distribution of time baselines and the number of observations of each star.

A succinct
overview of the APOGEE survey was presented in \citet{Eisenstein11}
and a full summary will be given by S.~Majewski et al. (2015, in
preparation). 
The APOGEE spectroscopic data processing is 
described in \citet{Nidever15}.
The pipeline for deriving atmospheric parameters and abundances from
the spectra will be described by A.~E.~Garc\'{\i}a~P\'erez (2015, in preparation).
The spectra, stellar parameters, and abundances for DR11 and DR12 are
described in \citet{Holtzman15}.

Figure~\ref{fig:APOGEE_DR12_logg_Teff_metals_cm_nm_alpham_nn2} shows the observed distribution
of the key stellar parameters and abundances for APOGEE DR12.
Obtaining robust and calibrated values of $T_{\rm eff}$, $\log{g}$, and
[M/H] along  
with individual abundances for
15 elements has required development of new stellar libraries
\citep{Zamora15} and $H$-band spectral line lists 
\citep{Shetrone15}.  After describing these
fits, we discuss a value-added catalog of red clump stars, then
describe specific target classes of APOGEE stars that are new since
DR10. 

\subsection{Abundances of 15 Elements in APOGEE DR12}

In DR12, we provide the best fitting values of the global stellar
parameters, as well as individual elemental abundances for C, N, O,
Na, Mg, Al, Si, S, K, Ca, Ti, V, Mn, Fe, and Ni.

The spectra are fit to models based on spectral libraries from
astronomical observations combined with laboratory and theoretical
transition probabilities and damping constants for individual
species. 
The final measurements and associated uncertainties are calibrated to
observations of stellar clusters, whose abundance patterns are assumed
to be uniform. 

The abundances are most reliable for stars with effective surface
temperatures in the range 3800~K$\leq T_{\rm eff}\leq$~5250~K.  
 For cooler atmospheres ($T_{\rm eff} < 3800$~K), the strengths of
 molecular transitions are increasingly sensitive to temperature, surface
 gravity, molecular equilibrium, and other physical details, giving
 rise to a greater uncertainty in the inferred abundances.   
 Stars with warmer atmospheres ($T_{\rm eff} > 5250$~K) or at low
 metallicity ([Fe/H]~$\lesssim -1$) have weaker lines, making it more
 difficult to measure abundances.

\subsection{Red Clump Stars in APOGEE}
This APOGEE data release also contains the DR11 and DR12 versions of
the APOGEE red-clump (APOGEE-RC) catalog. Red clump stars,
helium core-burning stars in metal-rich populations, are
good standard candles, and thus can be used as a spatial tracer of the
structure of the disk and the bulge.  RC stars are selected using the
$\log{g}$, [Fe/H], and near-infrared colors available for each APOGEE
star.  The construction of the DR11
APOGEE-RC catalog and the derivation of the distances to individual
stars were described in detail by \citet{Bovy14}.  
The DR11 catalog contains 10,341 stars with distances
accurate to about 5\%, with a contamination estimated to be
$\lesssim 7\%$. 

The DR12 RC catalog applies the same selection criteria to the full
DR12 APOGEE sample, but re-calibrates the surface gravities to a scale
appropriate for RC stars; the standard DR12
surface-gravity calibration is not appropriate for RC stars. The
calibration starting from the 
uncalibrated outputs of ASPCAP
for surface gravity, $\log g_{\mathrm{uncal.\,DR12}}$ is
$$
  \log g_{\mathrm{RC}} = 1.03\,\log g_{\mathrm{uncal.\,DR12}}-0.370\,,
$$
for $1 < \log g_{\mathrm{uncal.\,DR12}} < 3.8$ (outside of this range
the $\log g_{\mathrm{RC}}-\log g_{\mathrm{uncal.\,DR12}}$ correction is
fixed to that at the edges of this range). The DR12 APOGEE-RC catalog
contains 19,937 stars with an estimated contamination $\lesssim
3.5\%$ (estimated in the same manner as for the DR11 catalog, see
\citealt{Bovy14}).

\subsection{Additional Target Classes in APOGEE DR12}

Target selection for APOGEE was described in \citet{Zasowski13}.
As with BOSS, the targets for APOGEE are dominated by uniformly
selected samples designed to meet the key APOGEE science goals, but
also feature additional ancillary programs to take advantage of 
smaller-scale unique science opportunities presented by the APOGEE instrument.
The final distribution of 2MASS magnitudes and colors for 
all APOGEE targets are presented in
Figure~\ref{fig:APOGEE_DR12_star_color_dereddened_4panel}, both as
observed, and corrected for Galactic extinction.  Because many of the
APOGEE target fields are at quite low Galactic latitudes, the
extinction corrections can be quite substantial, even in the
infrared.  

Some of the additional dark time from the early completion of the BOSS
main survey was used for the existing APOGEE main program, and allowed the
addition and expansion of several ancillary science programs.  DR12
adds four additional ancillary target classes to those described in
\citet{Zasowski13} and extends two previous
ancillary programs.  We briefly describe these additions here:

{\bf Radial Velocity Monitoring of Stars in IC~348:}
The ``Infrared Spectroscopy of Young Nebulous Clusters'' (IN-SYNC)
ancillary program originally observed the Perseus sub-cluster IC~348.   
Subsequent to those observations a set of stars was targeted for 
further follow-up to 
 (1) search for sub-stellar companions in bright field stars of all spectral types;
 (2) search for stellar and sub-stellar companions around low-mass M stars;
 (3) search for pre-main-sequence spectroscopic binaries in IC~348;
 (4) study a newly identified Herbig Be object (HD 23478/BD+31 649)
 and
 (5) enhance the completeness of the IC~348 sample with 40 additional targets.
These 122 stars are labeled with {\tt APOGEE\_TARGET2} bit set to 18.

{\bf Probing Binarity, Elemental Abundances, and False Positives Among the Kepler Planet Hosts:}
This ancillary project observed 159 Kepler Objects of Interest (KOI;
e.g., \citealt{Burke14}),
23 M dwarfs, and 25 eclipsing binaries, at high cadence ($\sim$21
observations), over a period of 8 months to study binarity,
abundances, and false positives in the planet host sample.  
This project aims to detect stellar and brown dwarf companions of 
Kepler host stars, provide detailed
abundances for several elements, and understand planet formation in
binary systems.  KOI targets were selected from the KOI catalog with
$H_{Vega}<14$; ``eclipsing binary'' targets were selected with
$H<13$, periods $>5$~days, and 
classified as having a ``detached morphology'' as listed in the catalogs of \citet{Prsa11} and \citet{Slawson11}, plus two systems from \citet{Gaulme13};
and ``M dwarf'' targets were drawn from the catalog of \citet{Dressing13} with $T_{\rm eff} < 3500$~K and $H<14$. 
These 208 stars are labeled with {\tt APOGEE\_TARGET2} bit set to 19.

{\bf Calibration of the Gaia-ESO Spectroscopic Survey Program:}
A sample of 41 stars was observed to provide improved calibration of stellar parameters in conjunction with the Gaia-ESO Survey\footnote{\url{http://www.gaia-eso.eu/}} \citep{Pancino12}.
These observations are labeled with the setting of {\tt APOGEE\_TARGET2} bit 20.

{\bf Re-Observation of Commissioning Bulge Stars to Verify Radial Velocity Accuracy:}
A set of 48 stars in the bulge of the Milky Way that had originally been observed
during the early commissioning phase of the APOGEE instrument was
re-observed to provide a verification of the APOGEE radial velocity estimates.
These observations are labeled with the setting of {\tt APOGEE\_TARGET2} bit 21.

In addition, two previous ancillary programs were expanded in DR12.
The IN-SYNC ancillary program ({\tt APOGEE\_TARGET2}=13) to study
young stellar objects in the Perseus molecular cloud (see
\citealt{Cottaar14} and \citealt{Foster15} for more details)
was expanded in DR12 to observe 2,634 stars in the Orion~A molecular cloud.
The APOGEE ancillary program to observe {\it Kepler} stars for
asteroseismology and stellar parameter calibration ({\tt
  APOGEE\_TARGET1}=27) proved extraordinarily useful (e.g.,
\citealt{Epstein14}) and was folded
into the main APOGEE target selection for DR12.

\section{Data Distribution}
\label{sec:distribution}

The data for DR11 and DR12 are distributed through the same
mechanisms available in DR10, with some URL modifications to
accommodate
the ongoing transition to \mbox{SDSS-IV} and an associated unification of
the SDSS web presence under the \verb'sdss.org' domain.
Raw and processed image and spectroscopic data are available through
the Science Archive
Server\footnote{\url{http://data.sdss3.org/sas/dr12}}~\citep{Neilsen2008}
and through an interactive web
application.\footnote{\url{http://dr12.sdss3.org}}  
 The catalogs of photometric, spectroscopic, and derived quantities
are available through the Catalog Archive
Server\footnote{\url{http://skyserver.sdss.org/dr12}}~\citep{Thakar2008b,Thakar2008c}.   
More advanced and extensive querying capabilities are available
through ``CasJobs'', which allows time-consuming queries to be run in
the
background\footnote{\url{http://skyserver.sdss.org/casjobs}}~\citep{Li2008}.
GUI-driven queries of the database are
also available through
SkyServer.\footnote{\url{http://skyserver.sdss.org}}  
 Links to all of these methods are provided at
 \url{http://www.sdss.org/dr12/data\_access}.
The data processing software for APOGEE, BOSS, and SEGUE are publicly
available at \url{http://www.sdss.org/dr12/software/products}.
A set of tutorial examples for accessing SDSS data is provided at 
 \url{http://www.sdss.org/dr12/tutorials}.

\section{The Future: SDSS-IV}
\label{sec:future}

\mbox{SDSS-IV} began in 2014 July, as \mbox{SDSS-III} completed its observations.
It will continue the
legacy of SDSS with three programs on the 2.5-m Sloan Foundation Telescope to
further our understanding of our Galaxy, nearby galaxies, and the
distant Universe.

 The extended Baryon Oscillation Spectroscopic Survey 
  (eBOSS; K.~Dawson et al.~2015, in preparation)
  is obtaining spectra of LRGs over the redshift range $0.6<z < 1.0$
  and quasars in the range $0.9<z<3.5$ over 7500~deg$^2$,  
  and ELGs from $0.6<z<1.0$ over 1500~deg$^2$, 
  with an aim to measure the BAO peak to an
 accuracy of $<2$\% in four redshift bins.  eBOSS
 also includes a spectroscopic survey (TDSS) of
 variable stars and quasars (the Time Domain Spectroscopic Survey; TDSS; E.~Morganson et al.~2015, in preparation), along with
 a program to obtain optical spectra of 
 X-ray selected sources (The SPectroscopic IDentification of ERosita
 Sources; SPIDERS).  Many of the BOSS ancillary programs described in
 Appendix~\ref{appendix:boss_ancillary} are exploratory or pilot
 studies to test aspects of eBOSS target selection.  

 \mbox{SDSS-I/II} established our understanding of galaxies in the $z\sim0.1$ Universe. 
 The \mbox{SDSS-IV} Mapping Nearby Galaxies at APO (MaNGA) program
 \citep{Bundy15} will revisit 10,000 of these galaxies in far greater
 detail using integral-field fiber bundles to study spatially-resolved
 galaxy properties, star formation, and evolution. 
 
As Figure~\ref{fig:APOGEE_DR12_sky_coverage} makes clear, APOGEE has
sampled only a fraction of the Milky Way, and has missed the
Southern skies completely.  The APOGEE exploration of the Milky Way
will continue with SDSS-IV. APOGEE-2 will use the existing
spectrograph on the 2.5-m Sloan Foundation Telescope.  In addition, a second
APOGEE instrument will be built and installed on the 2.5-m du Pont
Telescope at Las Campanas Observatory, Chile, providing an all-sky view of
the Galaxy. 

  \mbox{SDSS-IV} will continue the sequence of SDSS public data releases,
  starting with a first release of spectroscopic data in 2016.

Data Release 12 marks the final data release of the \mbox{SDSS-III} project,
which began development in 2006 and conducted six years of
fully-dedicated operations at APO.  In total, \mbox{SDSS-III} collected
2350 deg$^2$ of $ugriz$ imaging and about 3.4 million spectra.  The
total SDSS data set now contains over 5 million spectra, with
connections to nearly all areas of astrophysics.  The median
extra-galactic redshift is now 0.5.
We thank the full
\mbox{SDSS-III} collaboration and partner institutions for their tremendous
efforts toward the realization of the ambitious goals of the project,
and we look forward to the many public uses of this vast legacy
data set.

\acknowledgments

\mbox{SDSS-III} Data Release 12 has made use of data products from the
Two Micron All Sky Survey, which is a joint project of the University
of Massachusetts and the Infrared Processing and Analysis
Center/California Institute of Technology, funded by the National
Aeronautics and Space Administration and the National Science
Foundation. 

\mbox{SDSS-III} Data Release 12 based APOGEE targeting decisions in
part on data collected by the Kepler mission. Funding for the Kepler mission is provided by the NASA Science Mission directorate.

\mbox{SDSS-III} Data Release 12 based MARVELS targeting decisions in part on the Guide Star Catalog 2.3.  The Guide Star \mbox{Catalogue-II} is a joint project of the Space Telescope Science Institute and the Osservatorio Astronomico di Torino. Space Telescope Science Institute is operated by the Association of Universities for Research in Astronomy, for the National Aeronautics and Space Administration under contract NAS5-26555. The participation of the Osservatorio Astronomico di Torino is supported by the Italian Council for Research in Astronomy.  Additional support is provided by European Southern Observatory, Space Telescope European Coordinating Facility, the International GEMINI project and the European Space Agency Astrophysics Division. 

\mbox{SDSS-III} Data Release 12 selected a significant number of BOSS ancillary targets based on data products from the Wide-field Infrared Survey Explorer, which is a joint project of the University of California, Los Angeles, and the Jet Propulsion Laboratory/California Institute of Technology, funded by the National Aeronautics and Space Administration.

\mbox{SDSS-III} Data Release 12 made use of Astropy, a community-developed core Python package for Astronomy \citep{AstroPy13}.

\mbox{SDSS-III} Data Release 12 made use of the Exoplanet Orbit Database and the Exoplanet Data Explorer at exoplanets.org.

\mbox{SDSS-III} Data Release 12 made use of the NASA/IPAC Extragalactic Database (NED) which is operated by the Jet Propulsion Laboratory, California Institute of Technology, under contract with the National Aeronautics and Space Administration. 

\mbox{SDSS-III} Data Release 12 made use of data from Pan-STARRS1.
The Pan-STARRS1 Surveys (PS1) have been 
made possible through contributions by the Institute for Astronomy, the 
University of Hawaii, the Pan-STARRS Project Office, the Max-Planck 
Society and its participating institutes, the Max Planck Institute for 
Astronomy, Heidelberg and the Max Planck Institute for Extraterrestrial 
Physics, Garching, The Johns Hopkins University, Durham University, 
the University of Edinburgh, the Queen's University Belfast, the 
Harvard-Smithsonian Center for Astrophysics, 
the Las Cumbres Observatory Global Telescope Network Incorporated, 
the National Central University of Taiwan, 
the Space Telescope Science Institute, and the National 
Aeronautics and Space Administration under Grant No. NNX08AR22G issued 
through the Planetary Science Division of the NASA Science Mission 
Directorate, the National Science Foundation Grant No. AST-1238877,
the University of Maryland, E\"otv\"os Lor\'and University (ELTE),
and the Los Alamos National Laboratory. 

Funding for \mbox{SDSS-III} has been provided by the Alfred P. Sloan Foundation,
 the Participating Institutions, the National Science Foundation, and the U.S. Department of Energy Office of Science. 
The \mbox{SDSS-III} web site is \url{http://www.sdss3.org/}.

\mbox{SDSS-III} is managed by the Astrophysical Research Consortium for the Participating Institutions of the \mbox{SDSS-III} Collaboration including the University of Arizona, the Brazilian Participation Group, Brookhaven National Laboratory, Carnegie Mellon University, University of Florida, the French Participation Group, the German Participation Group, Harvard University, the Instituto de Astrofisica de Canarias, the Michigan State/Notre Dame/JINA Participation Group, Johns Hopkins University, Lawrence Berkeley National Laboratory, Max Planck Institute for Astrophysics, Max Planck Institute for Extraterrestrial Physics, New Mexico State University, New York University, Ohio State University, Pennsylvania State University, University of Portsmouth, Princeton University, the Spanish Participation Group, University of Tokyo, University of Utah, Vanderbilt University, University of Virginia, University of Washington, and Yale University.

\bibliographystyle{apj}
\bibliography{refs,bits,inprep}

\newpage
\begin{appendix}
\section{Target Selection and Scientific Motivation for BOSS Ancillary Science Programs}
\label{appendix:boss_ancillary}

As described in \citet{Eisenstein11} and \citet{Dawson13}, up to 10\%
of the BOSS targets 
were reserved for ancillary programs, i.e., those with scientific aims
that went beyond those of the core quasar and galaxy samples.  
Ancillary programs observed in the 2009--2010 and 2010--2011 seasons
are documented in \citet{Dawson13}, and those observed in the
2011--2012 season were documented in \citet{DR10}. 
There were additional categories of ancillary programs included in the
2012--2014 observing seasons, which are released for the first time
with DR12, and which we document here.  In particular, BOSS completed
observations of its uniform galaxy and quasar samples over the full
footprint (Figure~\ref{fig:boss_coverage}) several months before the
end of SDSS-III observing, allowing a number of focused programs to be
carried out.  

All BOSS ancillary programs initiated after 2012 can be identified by
having a non-zero {\bf ANCILLARY\_TARGET2} bitmask.  We present in
this Appendix the scientific motivation for each program, the number
of fibers assigned, and a description of the target selection
algorithms.  The labels for each target bit name appear in bold font
in what follows. The new programs fall into three categories: those
that are dispersed throughout the remainder of the BOSS footprint at
low density (``parallel ancillary programs'',
Section~\ref{sec:boss_parallel}, Table~\ref{table:boss_parallel}), those
that were located in small regions of sky at high density (``dedicated
ancillary programs'', Section~\ref{sec:boss_dedicated},
Table~\ref{table:boss_dedicated}), and those associated with a pilot
survey in advance of eBOSS (``SEQUELS programs'',
Section~\ref{sec:boss_sequels}, Table~\ref{table:boss_sequels}).  Most
of the latter two categories were observed in the last six months of
SDSS-III observations, after the main survey had been completed.  Some
of these programs are self-contained science projects in themselves,
some represent calibrations or refinements of SDSS or BOSS
spectroscopic programs, and some, like the SEQUELS programs, are
preparatory for future surveys, especially eBOSS.  While few of these
programs have generated published results at this writing, a
significant number of papers are in preparation which use these data.
Note that there is often scientific 
or algorithmic overlap between many of the programs, reflecting the
multiple calls for ancillary programs within the SDSS collaboration.  


The selection algorithms in these different programs
typically use PSF, model (for galaxy magnitudes), or 
cmodel (for galaxy colors; \citealt{DR2}) SDSS photometry, all corrected
for Galactic extinction 
following \citet{SFD}.  Occasionally, fiber magnitudes are also
used.  The selection for many programs also uses photometry 
from the Wide-field Infrared Survey Explorer (WISE,
\citealt{Wright10}).
 WISE carried out a full-sky survey in four
bands, centered at 3.6, 4.5, 12, and 22$\mu$m; the resulting
photometry (which is reported on a Vega system, unlike the AB system
of SDSS) is referred to as W1, W2, W3, and W4, respectively, in what
follows.  We include the WISE catalog photometry of SDSS objects 
in both the DR11 and DR12 contexts in the CAS database.
A number of programs use a reprocessing of the WISE data
\citep{Lang14a} or forced photometry of WISE images at positions from SDSS
\citep{Lang14b}.

\subsection{BOSS Parallel Ancillary Programs}
\label{sec:boss_parallel}

All new parallel ancillary target classes found in DR12 were given a
priority lower than both the primary galaxy and quasar targets and
previously approved ancillary programs.  
The targeted samples for these parallel ancillary programs are therefore not
complete. 
We list
these programs roughly in the order of the distance to the targets; in
Table~\ref{table:boss_parallel}, we list them in bit order.  
\\

{\bf Characterizing Low-mass M Dwarfs Using Wide Binaries:}
M dwarf stars make up $\sim70\%$ of the stars in the Galaxy by number and have lifetimes longer than the age of the Universe.
They are thus valuable tracers of the chemical and dynamical evolution of
the Milky Way, but their complex spectra dominated by molecular bands
make it difficult to determine their ages and metallicities.  This
program targets earlier-type binary companions to known M dwarf
stars; these companions should share the same metallicity and age
as the M dwarf but have atmospheres that are easier to interpret.
These
systems can be used to refine relations between M dwarf properties and
spectral signatures \citep[e.g.,][]{Stassun08, Dhital12}. 


Fibers denoted by the {\bf SPOKE2} target flag were assigned to
candidate binary companions of spectroscopically confirmed low-mass
stars in the Sloan Low-mass Wide Pairs of Kinematically Equivalent Stars
\citep[SLoWPoKES;][]{Dhital10,Dhital15} project. A previous ancillary
program, the {\bf Low-Mass Binary Stars} program \citep{Dawson13}
consisted of systems with angular separations 65--180$\arcsec$.
{\bf SPOKE2} extends that
target sample to late-M spectral types, identifying binaries with
separations between 3 and 20 arcsec.  No proper motion requirement is
imposed  \citep{Dhital15}.
Targets have magnitudes in the range $17<i_{\rm PSF}<21.3$.
\\

{\bf A 
Census of Nearby Galaxies:}
We do not yet have a complete catalog of galaxies within 200 Mpc
\citep{Kasliwal11}, hampering studies of nearby transients and the
fine detail of the large-scale distribution of galaxies.
The Palomar Transient Factory (PTF; \citealt{Law09}) is performing a
narrow-band survey in two filters, centered at 656~nm and 663~nm, to
complete the catalog of galaxies in the local universe out to
200~Mpc. 
A sample of galaxies denoted by the {\bf PTF\_GAL} target flag was
selected for spectroscopic confirmation.  Galaxies without known
redshift were observed if they had an SDSS counterpart, a color
$m_{656} - m_{663} > 0.7$~mag, and relatively blue broadband color as
measured by SDSS ($g_{\rm model} - i_{\rm model} < 1.3$~mag).  Images
of all candidate galaxies were first visually inspected to avoid
spurious detections.  \\

{\bf Quasar Spectrophotometric Calibration:}
As described in
\citet{Dawson13} and \citet{Paris14}, the fibers assigned to BOSS CORE
and BONUS quasar targets \citep{Ross12} were offset in the focal plane
to optimize throughput in the blue part of the spectrum, to better observe
the Lyman-$\alpha$ forest. Because the standard stars are not observed
with this same offset, the spectrophotometric calibration of these
quasar targets is systematically incorrect.  The {\bf QSO\_STD} flag
denotes an additional sample of spectrophotometric standard stars,
from 10 to 25 per plate spread evenly across the focal plane, that
were drilled to follow the same offsets in the focal plane as the BOSS
quasar targets.  
These objects are chosen using the same algorithm as for normal
spectroscopic standard stars in BOSS, as explained in
\citet{Dawson13}.  Improved calibration gives improved measurements of
quasar spectral energy distributions, important both for constraining
quasar emission models, and for interpreting optical depth data in the
Lyman-$\alpha$ forest \citep{Lee15}. 
\\ 

{\bf Spectra of H$_2$O Maser Galaxies:}
One current route to the absolute calibration of the luminosities of Type Ia supernovae (SNeIa)
as standard candles uses the 3\%-accurate distance to 
NGC~4258 afforded by the well-studied H$_2$O maser in its center \citep{Humphreys13}. 
 Further improvements, by identifying other maser
galaxies with supernovae, could decrease the
uncertainty on local measurements of the Hubble Constant
\citep{Riess11}.  
There is an apparent correlation between maser activity and host
galaxy properties \citep{Zhu11}; this correlation will be tested with
spectroscopy of known maser host galaxies, and spectroscopy of SN~Ia host
galaxies will be used to identify plausible maser candidates.  Targets, identified with
the {\bf IAMASERS} flag, were selected with no previous SDSS spectra
and $i_{\rm model}< 20$~mag.  Objects targeted in the {\bf Bright
  Galaxies} ancillary program \citep{Dawson13} were also removed from
the {\bf IAMASERS} target list. 
\\

{\bf Spectroscopy of Massive Galaxy Cluster Members:}
This program aims to obtain redshifts of candidate
member galaxies of X-ray selected clusters.  
%
The sources are optical counterparts to X-ray clusters selected as faint
sources in the ROSAT All-Sky Survey \citep{Voges99,Voges00a}
identified by applying the redMaPPer \citep{Rykoff14} cluster finding
algorithm to the position of an X-ray source. 
The X-ray magnitude limit corresponds roughly to the brightest 30\% of
clusters that the X-ray satellite eROSITA will find within the BOSS area.

Objects denoted by the target flag {\bf CLUSTER\_MEMBER} are selected
from the redMaPPer catalog with $i_{\rm cmodel}<19.9$~mag
and $i_{\rm fib2}<21.5$~mag. Roughly 1000 candidate clusters were
observed. \\


{\bf Repeated Spectroscopy of Candidate Close Binary Massive Black Holes:}
Second-epoch spectroscopy was obtained for SDSS I/II quasars that are candidate massive black hole binaries with separations less than one parsec.
The quasars were selected from the DR7 quasar catalog \citep{DR7Q} based on having double-peaked broad Balmer lines or significant velocity offsets between broad
and narrow line centroids. The SDSS-III spectrum, identified by the {\bf DISKEMITTER\_REPEAT} target class, is separated from the first SDSS-I/II epoch by multiple
years and provides a test of binarity by observing changes in the emission line properties. These data should allow new constraints on the close massive
black hole binary population in SDSS quasars and will provide a better understanding of the nature of these peculiar broad line profiles.
\\

{\bf Spectroscopy of Hard X-ray Identified AGN:}
This sample, identified with the {\bf XMMSDSS} target class, was
designed to spectroscopically confirm hard ($2$--$10$~keV) X-ray selected
AGN identified in the serendipitous XMM survey of
SDSS~\citep{Georgakakis11}.  These objects tend to lie at relatively low
redshift, $z<0.8$.   
Objects identified by the {\bf XMMSDSS} target class were
selected with $f_X($2--10~keV$)>4\times10^{-14}$~\mbox{erg~cm$^{-2}$~s$^{-1}$}
 and SDSS $r_{\rm model}<22$~mag.  There was of order one target per
 square degree.  
The cross-correlation measurement of those AGN with the SDSS and BOSS
galaxy samples will constrain the dark matter halo masses of X-ray AGN as a
function of redshift and luminosity. \\

{\bf WISE BOSS:   BOSS spectra of Mid-IR bright AGN:}
Photometry from the WISE All-Sky Data Release
catalog was used in combination with SDSS photometry to select a 
12$\mu$m-flux-limited sample of quasars that goes beyond the main BOSS
quasar sample \citep{Ross12}. This allows studies of the completeness
of the main quasar sample and an exploration of dust obscuration of
quasars.  
The {\bf WISE\_BOSS\_QSO} target class was selected as having $i_{\rm PSF}<20.2$,
$\rm W1 -W2<0.30$, ${\rm W1} < 2.0+0.667 g_{\rm PSF}$, $r_{\rm
  PSF}-{\rm W2}> 2.0$, ${\rm W2}<18.5$, ${\rm W3}>12.5$,
and, for extended objects, $\rm W3>10.3$.
The resulting sample peaks at redshift $z \approx 1.4$.  
\\

{\bf Quasar Target Selection with WISE:}
This was a second sample of WISE-selected quasars, focused on the
redshift range $z > 2.15$.  
Candidate quasars, identified with the {\bf QSO\_WISE\_FULL\_SKY}
target class, were identified from SDSS photometry using an artificial neural network as described in \citet{Yeche10}.
Point sources are assigned a photometric redshift estimate and a likelihood ($NN$) ranging from zero
(stellar) to one (quasar). 
Objects with $NN>0.3$ were considered targets if they were matched within $1.5\arcsec$ of a WISE source,
had color $i_{\rm PSF}-{\rm W1} > 2.0 + 0.8 (g_{\rm PSF}-i_{\rm PSF})$
and $i_{\rm PSF}-{\rm W2}>3.0$, and were brighter than $g_{\rm
  PSF}=21.5$~mag. These color cuts were designed to identify
high-redshift quasars, and indeed almost 3/4 of the candidates have
redshifts above 2.  Objects satisfying this cut were assigned the {\bf
  QSO\_WISE\_FULL\_SKY} flag whether or not they were also targeted by the
main BOSS quasar selection. 
\\

{\bf Quasar Pairs:}
Candidate quasar pairs separated by angles corresponding to less than a few hundred kpc were
identified for spectroscopic confirmation. When combined with spectroscopy from other programs
\citep[e.g.,][]{Hennawi06,Myers08, Hennawi10}, this sample will provide a large statistical sample of
quasar pairs necessary for small-scale clustering measurements.
The target list consists of pairs of quasar targets selected using
either the Kernel Density Estimation (KDE) method \citep{Richards09}
or the XDQSOz method \citep{Bovy11}. There are both low- and
mid-redshift selection samples, both identified by the {\bf
  QSO\_XD\_KDE\_PAIR} target flag.  

The low-redshift selection includes targets with $g_{\rm PSF} < 20.85$~mag and a matching target from the same
selection within an angular separation, $\theta$, of $1\arcsec < \theta < 30\arcsec$.
Objects are selected based on being in the XDQSOz low-redshift selection
range ($0 < z < 2.2$) with probability being a quasar, $\rm PQSO > 0.8$;
or in the KDE catalog with flags indicating that the object is at low
redshift and/or has an ultraviolet excess ({\tt lowzts = 1} or {\tt
  uvxts = 1}, as described in Table 2 of \citealt{Richards09}).

The mid-redshift selection includes XDQSOz targets with $\rm PQSO > 0.2$
that have a pair (from the same mid-z selection) within 
$1\arcsec < \theta < 20\arcsec$.  These targets are further culled to only retain pairs for which the product of the two XDQSOz probabilities for the
pair integrated over $2.0 < z < 5.5$ is $\rm PQSO_1 \times PQSO_2 > 0.16$.

For both low- and mid-z selection, the following algorithm is
implemented to clean the sample: 
{\bf (1)} target all pairs where one or both of the objects in the
pair are in a BOSS tiling overlap region; 
{\bf (2)} for pairs where both objects are outside overlap regions, target the object with no existing spectrum;
{\bf (3)} for pairs where both objects are outside overlap regions and neither have existing spectra, target the fainter object;
{\bf (4)} discard all pairs where both objects are outside overlap regions and one of the pair is already a BOSS target;
{\bf (5)} discard all pairs where either object is a spectroscopically
confirmed star or is obviously an artifact on visual inspection
of the image; and
{\bf (6)} discard all targets (not pairs) that have an existing spectroscopic confirmation.
\\

\begin{deluxetable*}{llrrr}
\tablecaption{Parallel BOSS Ancillary Programs \label{table:boss_parallel}}
\tablehead{
\colhead{Primary Program}     & \colhead{Sub-Program} & \colhead{Bit Number} & \colhead{Number of Fibers\tablenotemark{a}} & \colhead{Number of Plates}
}
\startdata
QSO Selection with WISE       & QSO\_WISE\_FULL\_SKY &     10 &  26966 &    623   \\
Hard X-Ray AGN      & XMMSDSS              &     11 &     25 &     13   \\
H$_2$O Maser Galaxies            & IAMASERS             &     12 &     50 &     45   \\
Binary Black Holes            & DISKEMITTER\_REPEAT  &     13 &     92 &     70   \\
WISE BOSS                     & WISE\_BOSS\_QSO      &     14 &  20898 &    312   \\
Quasar Pairs & QSO\_XD\_KDE\_PAIR   &     15 &    628 &    273   \\
Galaxy Cluster Spectroscopy   & CLUSTER\_MEMBER      &     16 &   2757 &    268   \\
M Dwarf/Wide Binaries         & SPOKE2               &     17 &     93 &     65   \\
Census of Nearby Galaxies               & PTF\_GAL             &     19 &    173 &    107   \\
QSO Spectrophotometry         & QSO\_STD             &     20 &   1458 &    158   
\enddata
\tablenotetext{a}{More precisely, this is the number of spectra in
  each ancillary program that were denoted as ``specprimary'', i.e.,
  the best observation of a given object.  For ancillary programs that
  involved repeated observations of objects previously observed in
  BOSS, the number in this column may differ from the number of actual
  fibers drilled for the program by $<1$\%.} 
\end{deluxetable*}

\subsection{Ancillary Programs with Dedicated Plates}
\label{sec:boss_dedicated}

Because BOSS observations were proceeding ahead of schedule in 2012, a series
of plates were added to the SDSS-III program to observe ancillary science programs.
These plates do not have primary BOSS
galaxy and quasar targets and instead consist entirely of ancillary science targets.
The completeness of each dedicated sample is therefore typically higher than the completeness of the 
samples in the parallel ancillary programs.  We describe each of these
programs here, again sorted roughly by the distance of the targets.
Table~\ref{table:boss_dedicated} summarizes the target categories,
listed in order of {\bf ANCILLARY\_TARGET2} bit.  Note that a number
of the programs include multiple target classes, each indicated by a
separate bit. 
\\

{\bf Star Formation in the Orion and Taurus Molecular Clouds:}
This program obtained spectra of candidate young stellar objects (YSO)
in the Orion and Taurus molecular clouds.  The data provide a census
of YSO into the brown dwarf regime, a measurement of the initial mass
function at low masses, and a characterization of circumstellar disks
as a function of stellar mass, extending previous studies to fainter
magnitudes, to be sensitive to very low luminosity, low mass objects.
Objects were selected mostly from WISE photometry, as well as the Two
Micron All Sky Survey (2MASS; \citealt{Skrutskie06}) and {\it Spitzer}
photometry (matching to SDSS imaging where available;
\citealt{Finkbeiner04}) in the Orion and Taurus regions.  Objects were
included to the detection limit of the WISE catalog, but those with
$\rm W1>7$ were removed to reduce contamination from luminous, very
red, asymptotic giant branch stars.  There are several target classes
within this program, as detailed in Table~\ref{table:boss_dedicated}.

The five plates in this program
were designed in a heterogeneous manner
due to the different availability of SDSS imaging in each field and
the variation in the relative number of IR-excess sources.  The latter
is primarily related to the age of each star formation complex, as 
the circumstellar disk fraction decreases with stellar age. When limited SDSS
photometry is available in a field, $gri$ magnitudes are derived from
the PPMXL/USNO-B1 catalog following the inverse of the transformations
tabulated in \citet{Monet03}.

The 25 Ori spectroscopic plate targets WISE-detected stars within 1.5
degrees of the B3 star 25 Ori. It focuses on members of the young 25
Ori group and surrounding pre-main sequence stars in Orion and defines
the target classes {\bf 25ORI\_WISE} and {\bf 25ORI\_WISE\_W3}.

Objects were selected from the WISE catalog with detections in W1 and
W3, with a magnitude limit of $\rm W3 < 11.65$, and 
are assigned a target class of {\bf
  25ORI\_WISE\_W3}.  Sources were required to be fainter than 15 in
$g,r$ and $i$, and brighter than $g=22$ and $i=21$.  

The remaining three Orion plates covering the Kappa Ori, NGC 2023, and
NGC 2068 star formation regions were created in an identical manner
and define the target classes {\bf KOEKAP\_STAR}, {\bf KOEKAPBSTAR},
{\bf KOE2023\_STAR}, {\bf KOE2023BSTAR}, {\bf KOE2068\_STAR}, and {\bf
  KOE2068BSTAR}.  For all three plates, objects in the {\bf *\_STAR}
class are infrared excess sources selected by $W1-W2 > 0$ and a SNR in
W1 greater than 10. The {\bf *BSTAR} objects are other WISE detections within the field.

The Taurus spectroscopic plate targets objects with Spitzer
mid-infrared 8 and/or 24 micron excess within 1.5 degrees of the
center of the Taurus Heiles 2 molecular cloud. Our sample for Taurus
focuses on very low mass substellar objects with disks and edge-on
disks which may have been mistaken for galaxies.  The selection used
$\rm IRAC1 - IRAC4 > 1.5$ and/or $\rm IRAC1 - MIPS24 > 1.5$ mag with
$\rm SNR > 10$ for IRAC1 and $\rm SNR > 7$ for IRAC4 or MIPS24.  Here
IRAC1, IRAC4, and MIPS24 refer to Vega magnitudes measured through
filters centered at 3.5, 8.0, 
and 24 microns on {\it Spitzer}.  All science objects on the Taurus
plate have a target class of {\bf TAU\_STAR}.\\


{\bf Stars Across the SDSS:}
This project aims to cross-calibrate the large spectral surveys which
are giving us a detailed map of the different stellar populations in
the Milky Way.  Dedicated stellar spectroscopic surveys such as SEGUE
\citep{Yanny09}, the RAdial Velocity Experiment (RAVE;
\citealt{RAVE}), APOGEE, the Gaia/European Southern Observatory Survey
(GES; \citealt{Gilmore12}) and the massive $Gaia$ survey itself~\citep{deBruijne12} provide kinematic information and chemical diagnostics
for large samples of stars.  In addition, there are over 250,000 BOSS
spectra of stars (Table~\ref{table:dr12_contents}), mostly targeted as
quasar candidates.  Derived stellar parameters, such as effective
temperatures, surface graft's, and metallicities must be robust and
consistent between surveys to use them jointly to build a coherent
picture of our Galaxy.  Because each survey targets a particular
magnitude range, one must be careful to minimize systematic errors in
stellar parameters as a function of distance. 

This program obtained BOSS spectra of stars observed by the SEGUE-1
and SEGUE-2 surveys on eight plates (target classes {\bf SEGUE1}, {\bf
SEGUE2}), GES targets in eight plates (target class {\bf GES}), and
one plate dedicated to stars from the COnvection, ROtation,
and planetary Transit mission (CoRoT; \citealt{Corot}) also observed by GES and APOGEE
(target classes {\bf COROTGES}, {\bf COROTGETAPOG}).  As many CoRoT
and GES stars were given fibers as possible, restricted only with the bright
magnitude limit of $i>14$ to avoid saturation in the spectrographs.
There were not enough targets to fill all the fibers on the BOSS
plates, particularly when the GES fields did not overlap with SEGUE-1
or SEGUE-2 plates, so the eight {\bf GES} plates also targeted stars
selected from the SDSS photometry ({\bf SDSSFILLER}) with the
following selection cuts to ensure good SNR and to avoid
very cool stars for which it is more difficult to obtain accurate
stellar parameters with the SSPP: $0<g-r<1.25, g<19, i>15, r>15$.  The
CoRoT plate had targets chosen from APOGEE (target class {\bf APOGEE})
and 2MASS ({\bf 2MASSFILL}) as well.  Stars were targeted to sample
the full parameter space of effective temperature, metallicity, and
$\log g$, as much as possible.  The GES project (Milky Way survey)
targeted stars with $0< J - K <0.7, 12.5< J <17.5$, with near-infrared
photometry from the Visible and Infrared Survey Telescope for
Astronomy (VISTA; \citealt{VISTA}). In total, the eight GES plates gave
spectra with high enough SNR for acceptable SSPP parameters for 296
stars with $-0.25 < g-r<1.5$ and  $14 < g< 19$. 
\\


{\bf A Galaxy Sample Free of Fiber Collisions:}
The finite size of the BOSS fiber ferrules means that no two fibers
can be placed closer than $62''$ apart on a given plate.  These
``fiber collisions'' affect measurements of the small-scale
clustering of galaxies from the CMASS and LOWZ samples.  
CMASS and LOWZ galaxies that were not observed in the main BOSS survey due to
fiber collisions with other primary targets were added to ancillary
target plates 6373--6398 (North Galactic Cap), 6780--6782 (on Stripe 82),
 6369 and 6717. Fibers were also 
assigned to CMASS and LOWZ targets that suffered redshift failures ({\bf
ZWARNING\_NOQSO}$>$0; \citealt{Bolton12}) in previous observations in
the data reduction pipeline.  These objects are identified with the
CMASS or LOWZ target flags in the database; unlike all other objects
discussed in this Appendix, they are not assigned a target class in
{\bf ANCILLARY\_TARGET2}.  This program significantly increases the 
completeness of these galaxy samples in the region covered by these
plates and 
provides a useful dataset for testing the fiber-collision correction methods
that are currently used in BOSS clustering analyses (e.g.,
\citealt{Guo12}).  A
total of 1282 targets were included in this program.  These data have
been used in an analysis of velocity bias in close pairs of galaxies
by \citet{Guo15}.\\

{\bf Quantifying BOSS Galaxy Incompleteness with a WISE-Selected Sample:}
The CMASS sample is designed to select red galaxies of high stellar
mass ($M_{stellar} > 10^{11}\,M_\odot$).  This program (target class
{\bf WISE\_COMPLETE}) aimed to
explore a broader range of galaxy colors 
in the CMASS redshift range ($0.45 < z < 0.7$), using optical-IR cuts by combining SDSS
and WISE.  
%
The sample
criteria are $17.5 < i < 19.9$, $(r - ${\rm W1}$) > 4.165$,
and $i_{\rm fib2} < 21.7$ (the latter uncorrected for Galactic
extinction).  Various quality flag 
cuts were imposed to limit spurious sources.
Stars were eliminated using the SDSS morphological classifications 
for blue objects and a color-color cut in ($r-i$, $r-{\rm W1}$) space
for red objects. 
 A random subsample of 90\% of these objects were selected as targets to
meet the required target sky density.
\\

{\bf Exploring $z>0.6$ LRGs from SDSS and WISE:}
WISE and SDSS photometry was used to identify a sample of $z>0.6$
luminous red galaxies, taking advantage of the fact that the 1.6$\mu$m
bump in old stellar populations (due to a local minimum in the opacity
of the H$^-$ ion) is redshifted into the WISE W1 band.  This
spectroscopic sample will be used to calibrate photometric redshifts
in this range and to test target selection techniques for eBOSS.  

Targets for this program were divided into a higher priority sample
denoted {\bf HIZ\_LRG} and a lower priority sample denoted {\bf
  LRG\_ROUND3}.  
All objects were required to have 
$$(i_{\rm model} < 20.0\ ||\ z_{\rm model} < 20.0)\ \&\&\ (z_{\rm fib2} <
21.7\ ||\ i_{\rm fib2} < 22.0)\, .$$
Objects in the {\bf HIZ\_LRG} sample were selected to have 
$$(r - i) > 0.98\,\, \&\&\,\, (r - {\rm W1}) > 2 (r-i) -0.5.$$
The {\bf LRG\_ROUND3} sample used the same $r-\rm W1$ cut, but
the $(r-i)$ color cut was bluer, $(r - i) > 0.85$, in order to
explore a broader range of galaxy colors.  \\

{\bf Tests of eBOSS Target Selection in CFHTLS W3 Field:}
As a test of target selection algorithms to be used in eBOSS, six plates were dedicated to 
a selection of LRG and quasars at high density over a region of sky overlapping
the Canada-France-Hawaii Telescope Legacy Survey
(CFHTLS\footnote{http://www.cfht.hawaii.edu/Science/CFHTLS/}) W3
imaging footprint. 

Targets selected as potential galaxies in the redshift range
$0.6<z<0.9$ were denoted {\bf FAINT\_HIZ\_LRG}.  These objects were
selected in a similar manner to the targets that were assigned the
{\bf HIZ\_LRG} flag described above, but at fainter magnitudes with a
new tuning of color cuts.  Targets were required to have 
$$20 < z < 20.5,\ (z_{\rm fib2} < 22.2\ ||\ i_{\rm fib2} < 22.5), \ (r - i) >
0.98, \ (r - {\rm W1}) > 2 (r-i).$$  

Quasar targets, assigned the {\bf
  QSO\_EBOSS\_W3\_ADM} target class, 
were selected from photometry from CHFTLS, SDSS, and WISE, and
variability data from PTF.  
Five selection techniques were
applied, and all assigned the same target bit\footnote{The bit numbers
  in what follows are encoded in the bitmask {\bf W3bitmask}, included
  in the file
  \url{http://faraday.uwyo.edu/$\sim$admyers/eBOSS/ancil-QSO-eBOSS-W3-ADM-dr8.fits}.
}.  These selection criteria
were as follows: 
\begin{itemize}
\item {\bf Bit 0: W3 color box selection}.  These objects 
were selected from the CFHTLS W3 co-added catalog available at the TeraPix CFHT
website.\footnote{\url{\mbox{http://T07.terapix.fr/T07/Wide/W3/Big-Merged/W3\_fusion\_sm2.cat}}}  
 The objects were restricted in CFHT magnitudes to $g < 22.8$.  Stars
were excised with the following color cuts (using CFHT photometry): 
$$(g - r) - 0.5 (u - g) < -0.2\ ||\ (g - r) + 0.7 (u - g) < 0.6.$$
The targets were required to be classified as point sources by SDSS
and to have SDSS $r$ magnitudes in the range $17 < r < 22$. 

\item {\bf Bit 1: SDSS XDQSOz selection}.  These objects were selected using the
  XDQSOz selection of \citet{Bovy12b} based on SDSS photometry. Point
  sources with $17 < r <  22$ were required to have an XDQSOz probability of
  being a quasar greater 
  than 0.2.  

\item {\bf Bit 2: SDSS-WISE selection}. This program used WISE forced
  photometry at SDSS source positions \citep{Lang14a,Lang14b}.
A stacked flux was created in SDSS $gri$ ($m_{opt}$; with a relative
($g$,$r$,$i$) weighting of ($1,0.8,0.6$)), 
and a stacked flux was created in WISE W1 and W2 ($m_{\rm wise}$; with
(W1, W2) relative weights of $(1,0.5)$). Objects were
selected with $17 < m_{\rm opt} < 22$, $(g - i) < 1.5$, and $m_{\rm
  opt} - m_{\rm wise} > (g-i) + 3.0$.  Extended sources were allowed;
the sample was restricted to sources with a difference between SDSS
PSF and model magnitudes less than 0.1.


\item {\bf Bit 3: CFHTLS Variability selection}. Using three years of
  repeated observation in the one square degree field D3 of CFHTLS, 
objects were selected based on the variability measured in their light curves.
Objects were selected on $\chi^2$ and structure function parameters $A$ and $\Gamma$ \citep{Palanque-Delabrouille11} averaged over the three bands $gri$.
Using colors $c1$ and $c3$ defined as in \citet{Fan99}:
$c1\equiv 0.95 (u-g) + 0.31 (g-r) + 0.11 (r-i)$ and
$c3\equiv -0.39 (u-g) + 0.79 (g-r) + 0.47 (r-i)$,
two selections were applied.  The first used only CFHT information, requiring 
$A>0.08$, $\chi^2 >10.0$, $\Gamma>0.3$, $c3<0.6 -0.33c1$, and $g<23.0$.
The second used both CFHT and SDSS, and required that
$A>0.08$, $\chi^2 >10.0$, $\Gamma>0.2$, $g<22.0$, and that the object
be classified as point-like by SDSS. 

\item {\bf Bit 4: PTF variability selection}.  Using light curves
computed from PTF $R$-band imaging linked to SDSS $r$ with a color
correction, quasar candidates were
again selected by variability. All structure function or color-term
parameters are defined as above (cf., Bit 3). The objects were required
to have  $A>0.05$, $\chi^2 >10.0$ and $\Gamma>0.1$. In addition, the
objects were limited to $g<22.5$ and had to pass either of the
following two criteria based on SDSS photometry: a color and magnitude
cut  $r>18$ and $c3<1.0 -0.33c1$, or a color and morphology cut
requiring the object to be classified as point-like by SDSS and to
have a probability of being a quasar greater than 0.1 according to the
XDQSO algorithm.
\end{itemize}

{\bf eBOSS ELG Target Selection with Deep Photometry:}
This program used deep photometric data to select ELG
candidates, to assess algorithms for eBOSS. 
Photometry extending
to fainter limits than SDSS was used to assess algorithms for
selection of Emission Line Galaxies (ELG) for spectroscopic
observations.  In particular, blue star-forming galaxies in the redshift range
$0.6<z<1.2$ were selected from the CFHTLS Wide W3
field photometric redshift catalogue
T0007\footnote{\url{http://www.cfht.hawaii.edu/Science/CFHTLS/}}
\citep{Ilbert06, Coupon09}.  Targets with the {\bf FAINT\_ELG} target
class were selected at a density of nearly 
400 objects per square degree, and three plates were observed centered
on the same position.  The sample was defined to 
help evaluate the completeness of the targeting sample
and redshift success rates near the faint end of the ELG target
population.  

Selected objects satisfied the constraints: 
$$20<g<22.8, -0.5<(g-i)<2 \hbox{ and }-0.5<(u-r)<0.7 (g-i)+0.1.$$
All photometry was based on CFHTLS MAG\_AUTO magnitudes on the AB
system.  Objects with known redshift were excluded.  
These data are described in \citet{Comparat15}, which measured the
evolution of the bright end of the
$\left[\mathrm{O\textrm{\textsc{ii}}}\right]\,$ emission line luminosity function.
\\

{\bf The TDSS/SPIDERS/eBOSS Pilot Survey:}
This program carried out pilot observations in two fields for two components of the
SDSS-IV eBOSS survey: TDSS and SPIDERS (Section~\ref{sec:future}).  
The first field encompasses the existing XMM-Newton Large Scale Survey
(XMM-LSS), deep multi-band CFHTLS field imaging, and a Pan-STARRS1
(PS1; \citealt{Kaiser02,Kaiser10})
medium deep survey field (MD01) with hundreds of epochs.  The 
second field is also a PS1 medium deep field (MD03) located
in the Lynx/IfA Deep 
Field.  Both fields have 3--4 times as many PS1 epochs as does SDSS Stripe
82 \citep{Annis14}, and PS1 continued monitoring these fields at the
time the BOSS spectroscopy of these plates was carried out.  There were five target selection algorithms on these
plates, as follows:

Objects with the {\bf TDSS\_PILOT} target class were selected from PS1
photometry calibrated as described in \citet{Schlafly12}.  Targets
were selected by variability within each of the $gri$ filters, with
the requirement of a median PS1 magnitude $17<{\rm mag}_x<20.5$ and at least
30 observed epochs within that filter.  Objects were required to be
point-like in SDSS, with the difference between PSF and model magnitude less than 0.05 in
each filter, and with no detectable proper motions.  
Lightcurves for 
objects that pass a variability threshold in at least one filter
following \citet{Kim2011} were visually inspected in all three filters. 
 We assign each object a priority based on the number
of passed criteria summed over filters, the source brightness, and
whether or not a BOSS spectrum already exists.  

Objects identified {\bf TDSS\_PILOT\_PM} were selected 
the same way, but this identifier marks objects with significant 
($>3\,\sigma$) total proper motion as measured by SDSS. 

Objects identified {\bf TDSS\_PILOT\_SNHOST} showed transient behavior
in extended objects in the PS1 medium deep photometry, as described in
\citet{Chornock13}.  

Objects identified {\bf SPIDERS\_PILOT} were selected as X-ray
sources with clear optical counterparts in SDSS DR8 imaging.  
The X-ray selection was performed on a
source catalog constructed from public XMM-Newton data in the XMM-LSS
area following the procedure described in \citet{Georgakakis11}.  The
sample was flux-limited in soft X-rays (0.5--2 keV) to the expected
limit of the eROSITA deep 
field survey ($\sim6 \times 10^{-15}$~\mbox{erg~cm$^{-2}$~s$^{-1}$}),
and were
required to have $17<r_{\rm PSF}<22.5$ and not to have been 
spectroscopically observed by BOSS as of DR9.  Objects with higher
soft X-ray flux were given higher priority in fiber assignment.  

Objects targeted by both the SPIDERS and TDSS algorithms were given
higher priority and were assigned the {\bf TDSS\_SPIDERS\_PILOT} target
class. 
\\

{\bf Follow-up spectroscopy of wide-area XMM fields:}
Like the SPIDERS program above, this program targeted X-ray-selected
AGN from the XMM-XXL field, now using the full range of sensitivity
from 0.5 to 10 keV.  
%
SDSS optical counterparts to X-ray sources were identified via the
maximum-likelihood method \citep{Georgakakis11}. The main spectroscopic target sample
was selected to have $f_{X}(0.5$--$10$~keV$) > 10^{-14}$~\mbox{erg~cm$^{-2}$~s$^{-1}$}
and $15<r<22.5$, where $r$ is the PSF magnitude in
the case of optical unresolved sources or the model
magnitude for resolved sources.
Targets in this sample are denoted {\bf XMM\_PRIME}.
Secondary targets are sources with
$f_{X}(0.5$--$10$~keV$) <  10^{-14}$~\mbox{erg~cm$^{-2}$~s$^{-1}$}
and $15<r<22.5$, or radio sources selected in either 325 or 610\,MHz from
the catalogue of \citet{Tasse08}.  These targets are denoted {\bf XMM\_SECOND}.
\\

{\bf Multi-Object Reverberation Mapping:}
The broad emission lines in AGN spectra can have flux variations
correlated with variation in the continuum, but with a time delay
interpreted as the mean light-travel time across the broad-line region.
Measuring this time delay (``reverberation mapping'') allows one to study
the structure and kinematics of the broad-line regions of AGN. 
849 spectroscopically confirmed quasars were observed over 30 epochs
to study the variability of this sample. 
The observations were scheduled with a cadence of four to five days, as
weather allowed, with a goal of five epochs per month between 2014
January and the end of 2014 June.  
The typical exposure times were 2 hours for this program, 
and thus the final data from this program 
comprise a 60-hour effective exposure time for the targets in this field.
The survey is described in \citet{Shen15a}.  

Previous spectroscopy of the PS1 Medium Deep Field MD07
$(\alpha,\delta)=(213.704^\circ, +53.083^\circ)$ provided redshifts of
roughly 1200 quasars in the redshift range $0<z<5$ over the area of a
single plate.  The sample was limited to quasars with $i<21.7$.
Lower-redshift quasars (whose time delay should be easier to measure)
were given higher priority, and are indicated with the {\bf
  RM\_TILE1} target class; essentially all of these were assigned a
fiber.  Higher-redshift targets ({\bf RM\_TILE2}) were tiled with the
remaining fibers. 

Three plates containing identical science targets were drilled at varying hour angle
to ensure that the field was visible for six months.  Each plate was given the normal
number of sky fibers (80) but was allocated a substantially larger number of standard star
fibers (70 rather than 20) to allow more rigorous tests of
spectrophotometric calibration.  
Early science results from these data include measurements of the
velocity dispersions of the host galaxies of low-redshift quasars
from the high SNR co-added spectra \citep{Shen15b}, rapid
trough variability in broad absorption line quasars \citep{Grier15},
and the structure functions and time delays of a number of quasars. 
\\

{\bf Variability-selected Quasars at $1<z<4$ to $g=22.5$:}
The {\bf QSO\_VAR\_LF} bit labels a target class designed for studies
of the quasar luminosity function to 
$g<22.5$. The sample is located in Stripe 82 at $36^\circ< \alpha
<42^\circ$ where multi-epoch SDSS photometry is available, thus
enabling 
a variability selection with the neural network presented in
\citet{Palanque-Delabrouille11}.  Targets with 
point-like morphology that passed a loose variability criterion were selected
(neural network threshold of 0.5, where 1/0 indicates a quasar-like/stellar-like light curve).
Extended sources which satisfied the color selection $c3 < 0.6 - 0.33
c1$, where $c1$ and $c3$ are linear combinations of SDSS 
$ugriz$ bands as defined in \citet{Fan99}, were targeted if they
passed a tighter variability criterion (threshold of 0.9). 
Note that targets previously spectroscopically identified as quasars
were not included in the sample and therefore do not have the {\bf
  QSO\_VAR\_LF} bit set, even if they pass the selection criteria for
this program.  
\\

{\bf Faint End of the Quasar Luminosity Function:}
Targets that have the {\bf QSO\_DEEP} bit set used the same
variability selection as for {\bf QSO\_VAR\_LF}, but were selected in
the range $22 < g < 23.5$ from SDSS Stripe 82 data.  
Slightly extended objects with $-0.15 < (r_{\rm PSF} - r_{\rm model})
< 0.15$  were selected to a neural network threshold of 0.9. 

Additional targets were included in the sample when they had a large
probability of being a quasar according to the KDE \citep{Richards09}.  Unresolved objects with 
$\rm KDE(1.0<z<2.2) \geq 0.999$ or slightly resolved objects with
$-0.05 < (r_{\rm PSF}-r_{\rm model}) < +0.05$ and $\rm KDE(z>2.2) \geq
0.985$ were included.  Targets 
previously spectroscopically identified as quasars were not included
in the sample and therefore do not have the {\bf QSO\_DEEP} bit set 
even if they pass the selection criteria.

Finally, a sample of candidate Lyman-Break Galaxies was selected in color-space
and assigned the {\bf LBG} bit. These targets are slightly extended
objects that lie in one of two color-box regions: $0 < (g-r) < 0.15$
\&\& $(u-g) > (g-r) + 0.2$, or $0 < (g-r) <1.0$ \&\& $(u-g)>(g-r)+1.25$.
\\

{\bf SDSS-III Observations of LOFAR Sources:}
This ancillary program
was intended to target radio sources identified in deep observations
of the ELAIS-N1 region by the Low Frequency Array (LOFAR;
\citealt{LOFAR}).  LOFAR observations were planned with the high-band
antenna (HBA: 110--250 MHz) for roughly ten hours over 9 deg$^2$ to
eventually reach an rms depth of 100 $\mu$Jy at 150 MHz.  Spectroscopic
confirmation of these sources will provide insight into the
nature of the LOFAR radio population and aid in the science
exploitation of new radio surveys.  The LOFAR ELAIS N1 region is
well-studied by optical surveys and contains deep Jansky Very Large
Array (JVLA) and Giant Metre-Wave Radio Telescope (GMRT) imaging data
near the center of the field.

The LOFAR sample goes considerably deeper near the center of the
spectroscopic field, concentrating the targets there and making it
impossible to assign sky fibers uniformly over the focal plane.
Instead, there were a large number of fiber bundles that did not
contain a sky fiber and the usual sky interpolation routine in the
automated BOSS reductions could not be applied to the four plates designed
for this program.  For these plates, the data reduction pipeline was
modified to apply a constant sky model across each spectrograph (i.e.,
fibers 1--500 and 501--1000, respectively).  This results in larger
sky residuals than the typical calibrated BOSS spectra.  With this in
mind, users of
these data should treat the automated redshift classification and
narrow emission lines with caution.

All LOFAR radio sources were matched to SDSS optical counterparts
found within 2\arcsec\ of the radio source position.  The SDSS
position was used for the fiber placement.  The target classes selected for
this program are as follows:

{\bf ELAIS\_N1\_LOFAR} targets were selected from a preliminary image
of the ELAIS-N1 HBA data (115 to 190 MHz) that reached an rms noise
level of 333 $\mu$Jy.  Approximately 800 sources were detected to a
threshold of $1650$~$\mu$Jy and an additional 400 sources were
detected to a threshold of $1000$~$\mu$Jy.  These sources are
distributed over a field of radius approximately three degrees for a
total surface area of roughly 30~deg$^{-2}$. 
 In addition, 387 fainter LOFAR sources that 
could be clearly identified by eye 
in the ELAIS-N1 field were targeted. 

{\bf ELAIS\_N1\_FIRST} sources lacked a detection by LOFAR but
appeared in the catalog of the Faint Images of the Radio Sky at Twenty
cm (FIRST) survey \citep{FIRST95}, and had an SDSS optical counterpart
with $r_{\rm model}<23.0$. Fibers were placed at the SDSS position. 

{\bf ELAIS\_N1\_GMRT\_GARN} sources were identified from deeper GMRT
data at 610 MHz (rms depth of 40--70 $\mu$Jy) from the \citet{Garn08}
source catalog.  These sources are expected to be dominated by AGN. 

{\bf ELAIS\_N1\_GMRT\_TAYLOR} targets were also selected from GMRT
data \citep{Taylor14}, which are even
deeper (rms depth of 10 $\mu$Jy) than that used 
in the {\bf ELAIS\_N1\_GMRT\_GARN} sample. 
The deep GMRT radio catalog includes 2800 sources over 1.2 deg$^2$.  
The positional accuracy from the radio data appears to be better than 0.5\arcsec.

{\bf ELAIS\_N1\_JVLA} sources were also selected to be much fainter than the other samples.
The deep JVLA radio catalogue includes 483 sources over 0.13 deg$^2$ at
an angular resolution of 2.5\arcsec\ and RMS 
noise of 1 $\mu$Jy \citep{Taylor14}. The positional accuracy is similar to the {\bf ELAIS\_N1\_GMRT\_TAYLOR} sample.
Both this sample and the {\bf ELAIS\_N1\_GMRT\_TAYLOR} sample 
should include a significant fraction of star-forming galaxies at $z<1$. 
\\

\begin{deluxetable*}{llrrr}
\tablecaption{BOSS Ancillary Programs with Dedicated Plates \label{table:boss_dedicated}}
\tablehead{
\colhead{Primary Program}       &    \colhead{Sub-Program} & \colhead{Bit Number} & \colhead{Number of Fibers\tablenotemark{a}} & \colhead{Plate ID}
}
\startdata
ELG with Deep Photometry        & FAINT\_ELG              &     18 &   2588 & 6931--6933   \\
LRGs from SDSS and WISE         & HIZ\_LRG                &     21 &   8291 & 6373--6398   \\
LRGs from SDSS and WISE         & LRG\_ROUND3             &     22 &   2543 & 6373--6398   \\
Galaxy Incompleteness with WISE & WISE\_COMPLETE          &     23 &   9144 & 6373--6398   \\
TDSS/SPIDERS/eBOSS Pilot Survey & TDSS\_PILOT             &     24 &    859 & 6369, 6783   \\
TDSS/SPIDERS/eBOSS Pilot Survey & SPIDERS\_PILOT          &     25 &    363 & 6369, 6783   \\
TDSS/SPIDERS/eBOSS Pilot Survey & TDSS\_SPIDERS\_PILOT    &     26 &    107 & 6369, 6783   \\
Variability-Selected Quasars    & QSO\_VAR\_LF            &     27 &   2401 & 6370, 6780--6782   \\
TDSS/SPIDERS/eBOSS Pilot Survey & TDSS\_PILOT\_PM         &     28 &    129 & 6783   \\
TDSS/SPIDERS/eBOSS Pilot Survey & TDSS\_PILOT\_SNHOST     &     29 &      7 & 6783   \\
eBOSS in CFHTLS                 & FAINT\_HIZ\_LRG         &     30 &    684 & 7027--7032   \\
eBOSS in CFHTLS                 & QSO\_EBOSS\_W3\_ADM     &     31 &   3517 & 7027--7032   \\
Wide-Area XMM fields            & XMM\_PRIME              &     32 &   2422 & 7235--7238   \\
Wide-Area XMM fields            & XMM\_SECOND             &     33 &    648 & 7235--7238   \\
SEQUELS ELG                     & SEQUELS\_ELG            & 34\tablenotemark{b} &  4884 & 7239--7243,7245--7248\\
Stars Across SDSS               & GES                     &     35 &    410 & 7330--7333, 7450--7453   \\
Stars Across SDSS               & SEGUE1                  &     36 &   5262 & 7253--7256, 7454--7457   \\
Stars Across SDSS               & SEGUE2                  &     37 &   2104 & 7253--7256, 7454--7457   \\
Stars Across SDSS               & SDSSFILLER              &     38 &   4710 & 7330--7333, 7450--7453   \\
SEQUELS ELG                     &SEQUELS\_ELG\_LOWP       & 39\tablenotemark{b} &  3170 & 7239--7243,7245--7248\\
Orion and Taurus                & 25ORI\_WISE             &     40 &    290 & 7261   \\
Orion and Taurus                & 25ORI\_WISE\_W3         &     41 &    484 & 7261   \\
Orion and Taurus                & KOEKAP\_STAR            &     42 &    252 & 7260   \\
Orion and Taurus                & KOE2023\_STAR           &     43 &    202 & 7259  \\
Orion and Taurus                & KOE2068\_STAR           &     44 &    276 & 7257   \\
Orion and Taurus                & KOE2023BSTAR            &     45 &    563 & 7259   \\
Orion and Taurus                & KOE2068BSTAR            &     46 &    602 & 7257   \\
Orion and Taurus                & KOEKAPBSTAR             &     47 &    542 & 7260   \\
Stars Across SDSS               & COROTGESAPOG            &     48 &      2 & 7258   \\
Stars Across SDSS               & COROTGES                &     49 &     47 & 7258   \\
Stars Across SDSS               & APOGEE                  &     50 &    145 & 7258  \\
Stars Across SDSS               & 2MASSFILL               &     51 &    324 & 7258    \\
Orion and Taurus                & TAU\_STAR               &     52 &    734 & 7262   \\
SEQUELS                         & SEQUELS\_TARGET         &     53 & \nodata\tablenotemark{c} &7277--7329, 7374--7429 \\
Reverberation Mapping\tablenotemark{d}           & RM\_TILE1               &     54 &    230 & 7338--7340   \\
Reverberation Mapping\tablenotemark{d}           & RM\_TILE2               &     55 &    619 & 7338--7340   \\
Faint Quasars                   & QSO\_DEEP               &     56 &   2484 & 7334--7337   \\
Faint Quasars                   & LBG                     &     57 &    168 & 7336--7337   \\
LOFAR Sources                   & ELAIS\_N1\_LOFAR        &     58 &    410 & 7562--7565   \\
LOFAR Sources                   & ELAIS\_N1\_FIRST        &     59 &    321 & 7562--7565   \\
LOFAR Sources                   & ELAIS\_N1\_GMRT\_GARN   &     60 &    356 & 7562--7565   \\
LOFAR Sources                   & ELAIS\_N1\_GMRT\_TAYLOR &     61 &   1019 & 7562--7565   \\
LOFAR Sources                   & ELAIS\_N1\_JVLA         &     62 &     56 & 7562--7565   
\enddata
\tablenotetext{a}{More precisely, this is the number of spectra in each
  ancillary program that were denoted as ``specprimary'', i.e., the
  best observation of a given object.  For ancillary programs that
  involved repeated observations of objects previously observed in
  BOSS, the number in this column may differ from the number of actual
  fibers drilled for the program by $<1$\%.} 
\tablenotetext{b}{These targets are part of the SEQUELS program,
  described in Section~\ref{sec:boss_sequels}.}
\tablenotetext{c}{SEQUELS targets are discussed in detail in
  Section~\ref{sec:boss_sequels}.}
\tablenotetext{d}{These objects were observed over 30 epochs.  All
  these objects have previous spectra, and thus none of these observations are
  designated as ``specprimary''.}
\end{deluxetable*}

\subsection{The Sloan Extended Quasar, ELG, and LRG Survey (SEQUELS)}
\label{sec:boss_sequels}


SEQUELS serves both as a pilot program for the eBOSS survey of
SDSS-IV and as a stand-alone science program within SDSS-III.  
SEQUELS also encompasses two SDSS-IV sub-programs to obtain spectra of
variability-selected objects and X-ray detected objects, which are
pilot studies for the TDSS and SPIDERS programs within eBOSS described
in Section~\ref{sec:future}. 

The main SEQUELS footprint lies in the North Galactic Cap. 
Targets were selected over
the region covering $120^\circ<\alpha<210^\circ$ and
$45^\circ<\delta<60^\circ$ within the nominal BOSS footprint,
but only 300~deg$^{2}$ of this area were observed.
The targets in the primary SEQUELS program have the {\bf SEQUELS\_TARGET} bit set in the {\bf ANCILLARY\_TARGET2} bitmask.
Plates that were drilled but not observed before DR12 will be observed as part of eBOSS. 

SEQUELS targets fell into four broad categories, which we describe in
detail below: (1) luminous red
galaxies (LRG), designed to extend the
BOSS CMASS redshift coverage, yielding a median redshift of $\sim
0.72$; (2) quasars both as direct tracers of the cosmic density field
at redshifts $0.9<z<2.2$, and as probes of the Lyman-$\alpha$ forest;
(3) X-ray targets as a SPIDERS precursor, and (4) variability-selected
targets as a TDSS precursor.  Several other target classes don't fall
neatly into any of these categories and are listed at the end.  

In addition, SEQUELS incorporated a pilot program to
identify high-redshift ELGs.  The ELG targets are listed with the {\bf
  ANCILLARY\_TARGET2} bitmask (Table~\ref{table:boss_dedicated}).
The bitmasks for all other SEQUELS programs are listed in
Table~\ref{table:boss_sequels} and are described in detail in what
follows.  Note that some of these bits (such as bit 0, {\bf
  DO\_NOT\_OBSERVE}) don't indicate programs per se, but 
rather give information about the target selection process. 

%
 
\subsubsection{LRGs in SEQUELS}
Target selection of LRGs in SEQUELS was designed to target massive
red galaxies at $z\gtrsim 0.6$, using a combination of SDSS imaging and
WISE photometry. The SDSS photometry (all model magnitudes corrected
for Milky Way extinction) uses a new set of calibrations using a
combination of PanSTARRS-1 \citep{Kaiser10} and SDSS stellar
photometry \citep{Finkbeiner14}. The residual systematics are reduced
from 1\% in $griz$ \citep{Padmanabhan08} to 0.9, 0.7, 0.7 and 0.8\% in
the $griz$ bands, respectively.
In addition, some poorly-constrained zero-points
with errors exceeding 3\% in the DR9 data are now significantly
improved.  This new photometry will be
included in a future data release.  The WISE photometry (now
converted to the AB system) is forced photometry on SDSS positions
\citep{Lang14b}.  


There are two target classes focused on LRG; roughly 1/3 of the
LRG objects are targeted by both.  Both classes are magnitude
limited to $z<19.95$ and $i>19.9$.  The bright limit ensures that
there is no overlap with the BOSS CMASS selection. 
Objects flagged {\bf LRG\_IZW} in the SEQUELS bitmask
satisfy  the color cuts $(i-z)>0.7$ and $(i - {\rm W1})>2.143(i-z)-0.2$.
Objects flagged {\bf LRG\_RIW} satisfy $(r-i) > 0.98, (r - {\rm W1}) >
2 (r-i)$, and $(i-z) > 0.625$; the latter cut pushes the sample to
higher redshift.  
 
\subsubsection{Quasars in SEQUELS}
The main sample of SEQUELS quasars is  assigned the {\bf
  QSO\_EBOSS\_CORE} target class and is designed to meet the eBOSS
sky density goal of
$\sim 70$
$0.9 < z < 2.2$ quasars deg$^{-2}$.  
The target selection makes no attempt to filter out higher-redshift quasars, so
objects from this sample will also be useful for 
Lyman-$\alpha$ forest
studies. Quasars in the CORE are selected by a combination of XDQSOz \citep{Bovy12b}
in the optical and a WISE-optical color cut, as detailed in
A. Myers et al. (2015, in preparation); see also the description of
bit 1 and 2 of the {\bf QSO\_EBOSS\_W3\_ADM} target class above. 
This sample (and all the SEQUELS
quasar candidates which follow, unless otherwise indicated) are
restricted to objects classified as point sources, with faint-end
magnitude cuts of $g < 22$ or $r < 22$.

We also selected quasars via their variability as measured by the
PTF; these are given the target class {\bf QSO\_PTF}.  This sample is
less uniformly selected, given the availability of multi-epoch PTF
imaging, but that is acceptable for Lyman-$\alpha$ forest studies.
These objects 
are limited in magnitude to $r > 19$ and $g < 22.5$.

Targets that have the {\bf QSO\_EBOSS\_KDE} bit
set in SEQUELS consisted of all objects from the KDE catalog
of \citet{Richards09} that had {\tt uvxts=1} set (indicating that they had a UV
excess, and thus were likely to be at $z \lesssim 2.2$) within that catalog. 
Only KDE objects that matched to a point source in the DR9 
or the custom SDSS photometry used to select SEQUELS targets were included.

The {\bf QSO\_EBOSS\_FIRST} bit indicates quasars that are targeted in
SEQUELS because there is an SDSS source within 1\arcsec\ of a
source in the  2013 June 05 version\footnote{\url{http://sundog.stsci.edu/first/catalogs/readme\_13jun05.html}}
of the FIRST point source catalog \citep{FIRST95}. 

  An object is flagged {\bf QSO\_BOSS\_TARGET} if it has been
  previously observed by BOSS and does {\em not} have either {\tt
  LITTLE\_COVERAGE} or {\tt UNPLUGGED} set in the {\tt ZWARNING}
bitmask \citep[see Table 3 of][]{Bolton12}.  Similarly, an object from
SDSS DR8 is flagged {\bf QSO\_SDSS\_TARGET} if it is included in the SDSS
DR8 spectroscopic database, and similarly has neither of those flags
set in {\tt ZWARNING}.  



We separately flagged those quasars with {\bf QSO\_KNOWN} whose spectra had been visually
confirmed, as listed in the SDSS 
sample used to define known objects in BOSS \citep[see][]{Ross12}, 
and a preliminary version of the DR12 BOSS quasar
catalog of I. Paris et.~al. (2015, in preparation).

As part of SEQUELS, we also re-observed a number of high-redshift ($z>2.15$) quasars
that had low SNR spectroscopy in SDSS DR7 or BOSS, to improve the
measurement of the Ly$\alpha$ forest.  

Objects flagged {\bf
  QSO\_REOBS} had $0.75\leq {\rm SNR/pixel} <3$ in BOSS.  This target
class also included objects which have a high probability of being
quasars based on their photometry, but had no signal in the BOSS
spectra because of dropped fibers or other problems.  

In the same spirit,  BOSS spectra of some objects are of low enough
quality that their classification as quasars, or measurements of their
redshifts, are uncertain 
upon visual inspection. Such objects are designated as {\tt QSO?} or {\tt
  QSO\_Z?} in the DR12 quasar catalog (I. Paris et.~al.~2015, in preparation).
 Those objects in the SEQUELS footprint are re-observed, and given the {\bf
  QSO\_BAD\_BOSS} target class.  
A preliminary, but close-to-final version of
the DR12 catalog was used to define this sample for SEQUELS.

We set a flag bit, {\bf DO\_NOT\_OBSERVE}, to indicate which
previously observed quasars should not be re-observed, even if they
were selected by one of the SEQUELS algorithms.  It is determined by
the following combination of target flags:  

$$
{\bf  (QSO\_KNOWN~ ||~QSO\_BOSS\_TARGET~||~QSO\_SDSS\_TARGET)} \\ {\bf ~\&\&~!(QSO\_BAD\_BOSS~||~QSO\_REOBS)} ~.
$$

SEQUELS targeted quasars were selected in both the DR9 imaging used for BOSS {\em and}
an updated DR12 imaging calibration intended for use in eBOSS targeting. 
The {\bf DR9\_CALIB\_TARGET} bit signifies quasars that were selected for
SEQUELS using the DR9 imaging calibrations instead of (or as well as) the updated DR12 imaging.


\subsubsection{SPIDERS targets within the SEQUELS program}

The goal of the SPIDERS program within eBOSS is to obtain SDSS
spectroscopy for large samples of X-ray 
selected AGN and member galaxies of X-ray selected clusters. Two
SPIDERS pilot programs were executed within SEQUELS using
pre-eROSITA X-ray survey data.

{\bf SPIDERS\_RASS\_AGN} targets are candidate AGN detected in the
ROSAT All Sky Survey (RASS). A parent sample of X-ray sources was
formed from the concatenation of all Bright and Faint RASS catalogue
\citep{Voges99,Voges00b} detections lying within the SEQUELS footprint.
Given the large RASS positional uncertainties, we determine the
most probable optical counterpart for each RASS source using a novel
Bayesian algorithm (M.~Salvato et al, in preparation), an extension of the
method introduced by \citet{Budavari08} applied to all SDSS
photometric objects with $17< r< 22$ within 1\arcmin\ of each RASS
detection.  The algorithm uses the positional offset between each
possible association, the positional errors, and the colors of the
sources, given priors from a sample of previously matched XMM-Newton
sources \citep{Georgakakis11}.  Identified sources which already had
SDSS/BOSS spectra, were associated with objects in the
\citet{Veron-Cetty10} catalogue of known AGN, or were associated with
bright stars from the Tycho-II catalogue \citep{Hog00}, were removed.

%
Objects of type {\bf SPIDERS\_RASS\_CLUS} are selected from the
RedMapper catalogue (\citealt{Rykoff14}) of cluster members with
$17.0<i_{\rm fiber2}<21.0$ that lie in the SEQUELS footprint. A
prioritization scheme penalizes lower richness clusters and favors
highly-ranked members in the photometric red sequences. We also
targeted 22 clusters
selected in XMM-Newton observations by the XCLASS-RedMapper survey
\citep{Sadibekova14,Clerc12} with richness (i.e., number of
candidate members) greater than 20.  The high-quality XMM-Newton data allows more
detailed characterization of the cluster mass once the spectroscopic redshift is known
(via, e.g., derivation of intra-cluster gas temperatures).  Moreover,
the identification of these objects as clusters is unambiguous given
their X-ray data, so no cut is made on optical richness.  


\subsubsection{TDSS targets within the SEQUELS program}

The TDSS program targeted variable objects matched between imaging in both Pan-STARRS1
and SDSS. There are two classes of TDSS targets: single-epoch
spectroscopy (SES) and few-epoch spectroscopy (FES). \\

\noindent {\bf Single-epoch spectroscopy}: These targets comprise the main body of TDSS targets and are
flagged with target class {\bf TDSS\_A}.

We match SDSS point sources with $16 < i_{\rm psf} < 21$ to the PS1 
 ``uberCal'' database of 2013 September, restricting to objects with
more than 10 detections across the PS1 $griz$ bands.  
We also eliminate sources with a $g < 22$ neighbor
within 5\arcsec\ or an $i < 12$ neighbor within 30\arcsec\ to avoid
problems with deblending issues.

To identify variables within this subsample, we use a three-dimensional
Kernel Density Estimator.  We train our algorithm on known variables,
using the Stripe 82 variable catalog from \citet{ivezic07} and require
that the amplitude of variation in the $g$, $r$ and $i$ bands be greater
than 0.1.  Our catalog of non-variables is taken from the
\citet{ivezic07} standards catalog.  We improve the purity of
the latter catalog by requiring that our non-variables have at least eight SDSS
observations in Stripe 82 and a reduced $\chi^2$ relative to a model
of no variability 
of less than 2 in the $g$, $r$ and $i$ bands. We require that variables,
standards and candidates have SDSS and PS1 magnitude errors of less
than 0.1 and at least two PS1 detections in three of the four bands in
common between PS1 and
SDSS bands ($g$, $r$, $i$ and $z$).

Across the 3--4 qualified bands (as described above), we use the median
PS1-SDSS magnitude difference (corrected photometrically so that it is
0 for a typical star), median PS1-only variability (essentially the
variance minus the average error squared) and median SDSS magnitude as the
three dimensions of our KDE.  We bin and convolve both our variable
and standard population within this space and define ``efficiency'' as
the fraction of variables divided by the fraction of standards in
every region of that space. We then use the PS1-SDSS difference, PS1
variability and median magnitude to assign an efficiency to every
source in our sample.

We limit ourselves to sources with SDSS $i_{\rm fib2}<21$, and
fainter than 17 in $u, g$ and $r$ fiber magnitudes.  
This removes
potentially saturated sources.  We also remove targets that already
have SDSS or BOSS spectroscopy. \\

\noindent {\bf Few-epoch spectroscopy}: These target bits represent FES programs that
explicitly seek repeat spectra for objects of interest in order to
monitor spectroscopic variability.  
 The TDSS\_FES program targets are: 

\begin{itemize}

\item[] {\bf TDSS\_FES\_DE}: Quasar disk emitters. These targets are
  quasars with $i<18.9$ and broad, double-peaked or asymmetric
  Balmer emission line profiles, such as those in \citet{strateva06}
  ($z<0.33$ for H$\alpha$ and H$\beta$) and higher-redshift analogs
  from \citet{luo13} ($z\sim 0.6$ for H$\beta$ and Mg\,II).  This
  program seeks to  characterize the variability of the broad emission line
  profiles, especially changes in asymmetry and velocity profiles, for
  comparison to models of accretion disk emission in the presence of
  asymmetries and/or perturbations.

\item[] {\bf TDSS\_FES\_DWARFC}: Dwarf carbon stars (dCs). Most
  targets were chosen from the compilation of \citet{green13} from
  SDSS spectroscopy.  Objects were required to have 
  significant (more than $3\,\sigma$) proper motion ($\approx
  15\,$mas/yr) between the
  Palomar Observatory Sky Survey and SDSS photometry, ensuring that
  they are nearby, and thus likely to be dwarf stars. 
Observations of radial velocity variations will identify binaries,
thus testing the hypothesis
that these stars became carbon-rich due to mass transfer from an
asymptotic branch star via either wind accretion or Roche lobe
overflow. 

\item[] {\bf TDSS\_FES\_NQHISN}: This program targets $z<0.8$ DR7
  quasars with high SNR spectra to study broad-line variability on
  multi-year timescales. 

\item[] {\bf TDSS\_FES\_MGII}: This program targets quasars that
  showed evidence for temporal velocity shifts in the Mg\,II broad
  emission lines in previous repeat SDSS spectroscopy \citep{Ju13}
  in order to
  look for evidence of super-massive black hole binaries. 

\item[] {\bf TDSS\_FES\_VARBAL}: 
These   objects are selected from the \citet{gibson08} broad
absorption line  quasar catalog, to look for variability in the
absorption troughs. 
Further description of this program can be found in
  \citet{FilizAk12,FilizAk13}.

\end{itemize}

\subsubsection{Other Target Classes in SEQUELS} 

Galaxies from the main BOSS target selection,
both LOWZ and CMASS, that were not assigned fibers due to fiber
collisions were observed in SEQUELS and given the target class {\bf
  SEQUELS\_COLLIDED}. Observing these galaxies in SEQUELS creates
large contiguous areas that have 100\% spectroscopic completeness in
the final BOSS data sample.  A similar sample was described in
Section~\ref{sec:boss_dedicated}.

Variable targets selected from the PTF survey are targeted with the
{\bf SEQUELS\_PTF\_VAR} target class in three classes: hosts of
supernovae detected in the PTF supernova program, RR Lyrae stars, and
additional sources whose light-curve built from PTF data show
variations by 0.4 magnitude or more. 


Emission-line galaxy candidates tend to have blue colors and thus are
relatively bright in the $u$ band.  The South Galactic Cap U-band Sky
Survey\footnote{\url{http://batc.bao.ac.cn/Uband/}} (SCUSS) was carried
out over the SEQUELS area using the 2.3-m 
Bok Telescope at Kitt Peak to obtain deeper data ($u \approx 23$ for
$5\,\sigma$ detections of point sources) than SDSS 
(X.~Zhou et al.~2015, in preparation;
  H.~Zou et al.~2015, in preparation). 
 We used these data together with
SDSS $g,r,i$ photometry to select ELGs in the redshift range $0.4 < z
< 1.6$  
in a region of the sky of
25.7 deg$^2$ around $(\alpha,\delta)\sim(23^\circ,20^\circ$).

The brightest and bluest galaxy population ({\bf SEQUELS\_ELG}) 
 is selected by
\begin{equation}
-0.5<u-r<0.7(g-i)+0.1 ~\&\&~20<u<22.5
\end{equation}
To fill the remaining fibers we
also observed targets satisfying broader color cuts ({\bf
  SEQUELS\_ELG\_LOWP}):  
\begin{equation}
(20<u<22.7 ~\&\&~ -0.9<u-r) ~\&\&~ (u-r<0.7(g-i)+0.2 ~||~ u-r<0.7)
\end{equation}

 

\begin{deluxetable*}{lrr}
\tablecaption{SEQUELS Targets \label{table:boss_sequels}}
\tablehead{
\colhead{Sub-Program} & \colhead{Bit Number} & \colhead{Number of Fibers\tablenotemark{a}}
}
\startdata
DO\_NOT\_OBSERVE       &  0                  &\nodata\tablenotemark{b}\\
LRG\_IZW               &  1                  & 11778 \\
LRG\_RIW               &  2                  & 11687 \\
QSO\_EBOSS\_CORE       & 10                  & 19461 \\
QSO\_PTF               & 11                  & 13232 \\
QSO\_REOBS             & 12                  &  1368 \\
QSO\_EBOSS\_KDE        & 13                  & 11843 \\
QSO\_EBOSS\_FIRST      & 14                  &   293 \\
QSO\_BAD\_BOSS         & 15                  &    59 \\
QSO\_BOSS\_TARGET      & 16                  & \nodata\tablenotemark{b} \\
QSO\_SDSS\_TARGET      & 17                  & \nodata\tablenotemark{b} \\
QSO\_KNOWN      & 18                  &        \nodata\tablenotemark{b} \\
DR9\_CALIB\_TARGET     & 19                  & 28602\tablenotemark{b} \\
SPIDERS\_RASS\_AGN     & 20                  &   162 \\
SPIDERS\_RASS\_CLUS    & 21                  &  1532 \\
TDSS\_A                & 30                  &  9418 \\
TDSS\_FES\_DE          & 31                  &    42 \\
TDSS\_FES\_DWARFC      & 32                  &    19 \\
TDSS\_FES\_NQHISN      & 33                  &    74 \\
TDSS\_FES\_MGII        & 34                  &     1 \\
TDSS\_FES\_VARBAL       & 35                  &    62 \\
SEQUELS\_PTF\_VAR & 40                  &   701 \\
SEQUELS\_COLLIDED      & 41                  &  \\
\enddata
\tablenotetext{a}{More precisely, this is the number of spectra in a ancillary program that were denoted as ``specprimary'', i.e., the best observation of a given object.  For ancillary programs that involved repeated observations of objects previously observed in BOSS, the number in this column may differ from the number of actual fibers drilled for the program by $<1$\%.}
\tablenotetext{b}{These bits are not target classes, but are
  identifiers of quasars targeted by other algorithms satisfying
  various criteria, as described in the text.}
\end{deluxetable*}

\end{appendix}

\end{document}